%% file: SUS-16-035_temp.tex
\pdfoutput=1

\documentclass[11pt,twoside,a4paper,cmspaper,final,collab]{cms-tdr}
\def\svnVersion{424047P}\def\cmsCernNoTag{CERN-EP-2017-071}\def\cmsCernDate{\today}\def\cmsMessage{Published in the European Physical Journal C as \href{http://dx.doi.org/10.1140/epjc/s10052-017-5079-z}{\doi{10.1140/epjc/s10052-017-5079-z.}}}

\begin{document}\cmsNoteHeader{SUS-16-035}

\hyphenation{had-ron-i-za-tion}
\hyphenation{cal-or-i-me-ter}
\hyphenation{de-vices}
\RCS$Revision: 405755 $
\RCS$HeadURL: svn+ssh://svn.cern.ch/reps/tdr2/papers/SUS-16-035/trunk/SUS-16-035.tex $
\RCS$Id: SUS-16-035.tex 405755 2017-05-21 19:51:39Z gzevi $
\newlength\cmsFigWidth
\ifthenelse{\boolean{cms@external}}{\setlength\cmsFigWidth{0.85\columnwidth}}{\setlength\cmsFigWidth{0.4\textwidth}}
\ifthenelse{\boolean{cms@external}}{\providecommand{\cmsLeft}{top\xspace}}{\providecommand{\cmsLeft}{left\xspace}}
\ifthenelse{\boolean{cms@external}}{\providecommand{\cmsRight}{bottom\xspace}}{\providecommand{\cmsRight}{right\xspace}}
\ifthenelse{\boolean{cms@external}}{\providecommand{\cmsTable}[1]{#1}}{\providecommand{\cmsTable}[1]{\resizebox{\textwidth}{!}{#1}}}

\providecommand{\NA}{\ensuremath{\text{---}}\xspace}
\newcommand{\sslumi}{35.9\fbinv}
\newcommand{\MTmin}{\ensuremath{m_\mathrm{T}^{\text{min}}}\xspace}
\newcommand{\Njets}{\ensuremath{N_\text{jets}}\xspace}
\newcommand{\Nbjets}{\ensuremath{N_{\PQb}}\xspace}
\newcommand{\gluino}{\PSg}
\newcommand{\susyq}{\PSQ}
\newcommand{\susytop}{\PSQt}
\newcommand{\sbottom}{\PSQb}
\newcommand{\lsp}{\PSGczDo}
\newcommand{\chiplus}{\PSGcpDo}
\newcommand{\chiplmin}{\PSGcpmDo}
\newcommand{\Totttt}{\ensuremath{\mathrm{T}1{\PQt\PQt\PQt\PQt}}\xspace}
\newcommand{\Tftttt}{\ensuremath{\mathrm{T}5{\PQt\PQt\PQt\PQt}}\xspace}
\newcommand{\Tfttcc}{\ensuremath{\mathrm{T}5{\PQt\PQt\PQc\PQc}}\xspace}
\newcommand{\TfqqqqWW}{\ensuremath{\mathrm{T}5{\PQq\PQq\PQq\PQq}\PW\PW}\xspace}
\newcommand{\TsttWW}{\ensuremath{\mathrm{T}6{\PQt\PQt}\PW\PW}\xspace}
\newcommand{\TfttbbWW}{\ensuremath{\mathrm{T}5{\PQt\PQt\PQb\PQb}\PW\PW}\xspace}
\newcommand{\MGAMCNLO} {{\MADGRAPH{}5\_a\MCATNLO}\xspace}

\cmsNoteHeader{SUS-16-035}
\title{Search for physics beyond the standard model  in events with two
  leptons of same sign, missing transverse momentum, and jets in
 proton-proton collisions at $\sqrt{s} = 13\TeV$}
\titlerunning{SUSY search with  two SS leptons, missing transverse momentum, and jets at 13\TeV}

\date{\today}

\abstract{
A data sample of events from proton-proton collisions
with two isolated same-sign leptons, missing transverse momentum,
and jets is studied in a search for signatures of new physics phenomena by
the CMS Collaboration at the LHC.
The data correspond to an integrated luminosity of 35.9\fbinv, and a center-of-mass
energy of 13\TeV.
The properties of the events are consistent with expectations from standard model processes,
and no excess yield is observed.
Exclusion limits at 95\% confidence level are set on cross sections for the pair
production of gluinos, squarks, and same-sign top quarks,
as well as top-quark associated production of a heavy scalar or
pseudoscalar boson decaying to top quarks, and on the standard model production of events with four top quarks.
The observed lower mass limits are as high as 1500\GeV for gluinos, 830\GeV for bottom squarks.
The excluded mass range for heavy (pseudo)scalar bosons is 350--360 (350--410)\GeV.
Additionally, model-independent limits in several topological regions
are provided, allowing  for further interpretations of the results.}

\hypersetup{%
pdfauthor={CMS Collaboration},%
pdftitle={Search for physics beyond the standard model  in events with two
  leptons of same sign, missing transverse momentum, and jets in
 proton-proton collisions at sqrt(s)=13 TeV},%
pdfsubject={CMS},%
pdfkeywords={CMS, physics, supersymmetry}}

\maketitle

\section{Introduction}
\label{sec:intro}
Final states with two leptons of  same charge, denoted as same-sign (SS) dileptons, are
produced rarely by standard model (SM) processes in proton-proton ($\Pp\Pp$) collisions.
Because the SM rates of SS dileptons are low, studies of these final states
provide excellent opportunities to search for manifestations of physics
beyond the standard model (BSM).  Over the last decades, a large number of new physics mechanisms
have been proposed to extend the SM and address its shortcomings.
Many of these can give rise to potentially large contributions
to the SS dilepton signature, e.g.,
the production of supersymmetric (SUSY)
particles~\cite{Barnett:1993ea,Guchait:1994zk}, SS top quarks~\cite{Bai:2008sk,Berger:2011ua},
scalar gluons (sgluons)~\cite{Plehn:2008ae,Calvet:2012rk}, heavy scalar bosons of extended Higgs sectors~\cite{Gaemers:1984sj,Branco:2011iw},
Majorana neutrinos~\cite{Almeida:1997em}, and vector-like quarks~\cite{Contino:2008hi}.

In the SUSY framework~\cite{Ramond:1971gb,Golfand:1971iw,Neveu:1971rx,Volkov:1972jx,Wess:1973kz,Wess:1974tw,Fayet:1974pd,Nilles:1983ge,Martin:1997ns,Farrar:1978xj},
the SS final state can appear in R-parity conserving models
through gluino or squark pair production when the decay of each of the pair-produced particles yields one or more \PW\ bosons.
For example, a pair of gluinos (which are Majorana particles) can give rise to SS charginos and up to
four top quarks, yielding signatures with up to four \PW\ bosons, as well as jets, \PQb quark jets, and large missing transverse momentum (\MET).
Similar signatures can also result from the pair production of bottom squarks, subsequently decaying to charginos and top quarks.

While R-parity conserving SUSY models often lead to signatures with large \MET,
it is also interesting to study final states without significant \MET beyond what is produced by the neutrinos from leptonic \PW\ boson decays.
For example, some SM and BSM scenarios can lead to the production of SS or multiple top quark pairs, such as the associated production of a heavy (pseudo)scalar, which subsequently decays to a pair of top quarks.
This scenario is realized in Type II two Higgs doublet models (2HDM) where
associated production with a single top quark or a \ttbar pair can in some cases provide a promising window
to probe these heavy (pseudo)scalar bosons~\cite{Dicus:1994bm,Craig:2015jba,Craig:2016ygr}.

This paper extends the search for new physics presented in Ref.~\cite{SUS-15-008}.
We consider final states with two leptons (electrons and muons) of  same charge,
two or more hadronic jets, and moderate \MET.
Compared to searches with zero or one lepton, this final state provides enhanced sensitivity to low-momentum
leptons and SUSY models with compressed mass spectra.
The results are based on an integrated luminosity corresponding to \sslumi
of $\sqrt{s} = 13\TeV$ proton-proton collisions collected with the CMS detector
at the CERN LHC.
Previous LHC searches in the SS dilepton channel have been performed by the
ATLAS~\cite{ATLAS:2012ai,Aad:2014pda,Aad:2016tuk}  and
CMS~\cite{SUS-10-004,SUS-11-020,SUS-11-010,SUS-12-017,SUS-13-013, SUS-15-008} Collaborations.
With respect to Ref.~\cite{SUS-15-008}, the event categorization is extended to take advantage of the increased integrated
luminosity, the estimate of rare SM backgrounds is improved, and the (pseudo)scalar boson interpretation is added.

The results of the search are interpreted in a number of specific BSM models
discussed in Section~\ref{sec:samples}.  In addition,
model-independent results are also provided in several kinematic regions to allow for further interpretations.
These results are given as a function of hadronic activity and of \MET, as well as in a set of inclusive regions with different topologies.
The full analysis results are also summarized in a smaller set of exclusive regions
to be used in combination with the background correlation matrix to facilitate their reinterpretation.

\section{Background and signal simulation}
\label{sec:samples}

{\tolerance=1200
Monte Carlo (MC) simulations are used to estimate SM background contributions and to estimate
the acceptance of the event selection for BSM models.
The \MGAMCNLO 2.2.2~\cite{MADGRAPH5,Alwall:2007fs,Frederix:2012ps} and
\POWHEG~v2~\cite{Melia:2011tj,Nason:2013ydw}  next-to-leading order (NLO) generators
are used to simulate almost all SM background processes
based on the NNPDF3.0 NLO~\cite{Ball:2014uwa} parton distribution functions (PDFs).
New physics signal samples, as well as the same-sign $\PW^{\pm} \PW^{\pm}$ process,
are generated with \MGAMCNLO at leading order (LO) precision, with up to two additional partons in the matrix element calculations,
 using the NNPDF3.0 LO~\cite{Ball:2014uwa} PDFs.
Parton showering and hadronization, as well as the double-parton scattering production of $\PW^{\pm} \PW^{\pm}$,
are described using the \PYTHIA~8.205 generator~\cite{Sjostrand:2007gs}
with the CUETP8M1 tune~\cite{Skands:2014pea,CMS-PAS-GEN-14-001}.
The \GEANTfour package~\cite{Geant} is used to model the CMS detector response for background samples,
while the CMS fast simulation package~\cite{Abdullin:2011zz} is used for signal samples.
\par}

To improve on the {\MADGRAPH} modeling of the
multiplicity of additional jets from initial-state radiation (ISR),
{\MADGRAPH} \ttbar MC events are reweighted based on the
number of ISR jets ($N_J^\mathrm{ISR}$), so as to make the light-flavor jet
multiplicity in dilepton \ttbar events agree with the one observed in data.
The same reweighting procedure is applied to SUSY MC events.
The reweighting factors vary between 0.92 and 0.51 for
$N_J^\mathrm{ISR}$ between 1 and 6.  We take one half of the deviation
from unity as the systematic uncertainty in these reweighting factors.

\begin{figure*}
\centering
\includegraphics[width=1\textwidth]{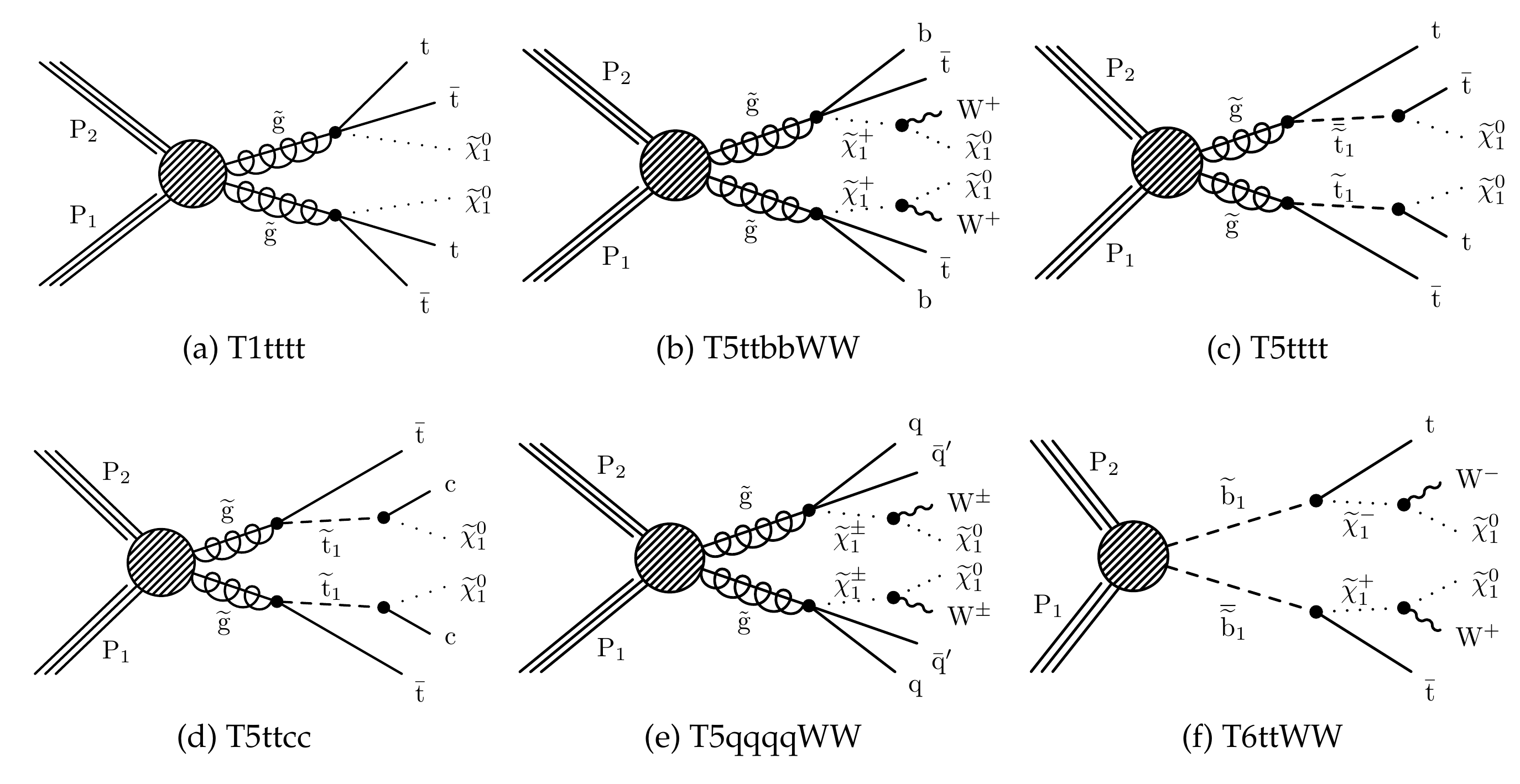}
\caption{Diagrams illustrating the simplified SUSY models considered in this analysis.}
\label{fig:diagrams}
\end{figure*}

\begin{figure*}
  \centering
  \includegraphics[width=1\textwidth]{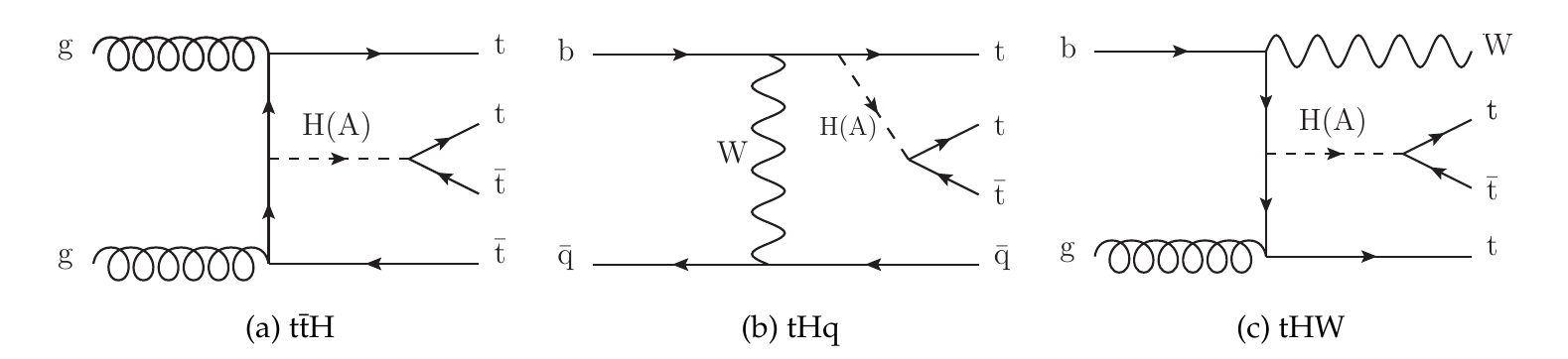}
    \caption{
    \label{fig:scalar_diagrams}
      Diagrams for scalar (pseudoscalar) boson production in association with top quarks. }
\end{figure*}

The new physics signal models probed by this search are shown in Figs.~\ref{fig:diagrams} and~\ref{fig:scalar_diagrams}.
In each of the simplified SUSY models~\cite{Alves:2011wf,Chatrchyan:2013sza} of Fig.~\ref{fig:diagrams},
only two or three new particles have masses sufficiently low to be produced on-shell,
and the branching fraction for the decays shown are assumed to be 100\%.
Gluino pair production models giving rise to signatures with up to four \PQb quarks and up to four \PW\ bosons are
shown in Figs.~\ref{fig:diagrams}a--e.
In these models, the gluino decays to the lightest squark ($\gluino \to \susyq \PQq$),
which in turn decays to same-flavor ($\susyq \to \PQq \lsp$)
or different-flavor ($\susyq \to \PQq' \chiplmin$) quarks.
The chargino decays to a \PW\ boson and a neutralino ($\chiplmin \to \PW^{\pm} \lsp$),
where the \lsp escapes detection and is taken to be the lightest SUSY particle (LSP).
The first two scenarios considered in Figs.~\ref{fig:diagrams}a and~\ref{fig:diagrams}b
include an off-shell third-generation squark ($\susytop$ or $\sbottom$) leading to the three-body decay of the gluino,
$\gluino \to \ttbar\lsp$ (\Totttt )
and $\gluino \to \cPaqt \PQb \chiplus$ (\TfttbbWW ),
resulting in events with four \PW\ bosons and four \PQb quarks.
In the \TfttbbWW model, the mass splitting between chargino and neutralino
is set to $m_{\chiplmin} - m_{\lsp} = 5\GeV$, so that two of the \PW\ bosons are produced off-shell and
can give rise to low transverse momentum (\pt) leptons.
The next two models shown (Figs.~\ref{fig:diagrams}c and d)
include an on-shell top squark with different mass splitting between
the $\susytop$ and the \lsp, and consequently different decay modes:
 in the \Tftttt model the mass splitting is equal to the top quark mass ($m_{\susytop} - m_{\lsp} = m_{\PQt}$),
 favoring the $\susytop \to \PQt \lsp$ decay,
 while in the \Tfttcc model the mass splitting is only $20\GeV$, favoring the flavor changing neutral current
  $\susytop \to \PQc \lsp$ decay.
In Fig.~\ref{fig:diagrams}e, the decay proceeds through a virtual light-flavor squark,
leading to a three-body decay to $\gluino \to \PQq \PQq' \chiplmin$, resulting in a signature
with two \PW\ bosons and four light-flavor jets. The two \PW\ bosons can have the same charge, giving rise
to SS dileptons. This model, \TfqqqqWW, is studied as a function of the gluino and
\lsp mass, with two different assumptions for the chargino mass: $m_{\chiplmin} = 0.5(m_{\gluino} +  m_{\lsp})$,
producing mostly on-shell \PW\ bosons,
and $m_{\chiplmin} = m_{\lsp}+20\GeV$, producing off-shell \PW\ bosons.
Finally, Fig.~\ref{fig:diagrams}f shows a model of bottom squark production followed by
the $\sbottom \to \PQt \chiplmin$ decay, resulting in two \PQb quarks and four \PW\ bosons.
This model, \TsttWW, is studied as a function of the \sbottom and \chiplmin masses,
keeping the \lsp mass at 50\GeV, resulting in two of the \PW\ bosons being produced off-shell
when the  \chiplmin and \lsp masses are close.
The production cross sections for SUSY models are calculated at NLO plus next-to-leading logarithmic (NLL)
accuracy~\cite{bib-nlo-nll-01,bib-nlo-nll-02,bib-nlo-nll-03,bib-nlo-nll-04,bib-nlo-nll-05,Borschensky:2014cia}.

The processes
shown in Fig.~\ref{fig:scalar_diagrams},
$\ttbar \PH$, $\PQt \PH \PQq$, and $\PQt \PW \PH$, represent the
top quark associated production of a scalar ($\PH$) or a pseudoscalar (\PSA).
The subsequent decay of the (pseudo)scalar to a pair of top quarks then gives rise to
final states including a total of three or four top quarks.
For the purpose of interpretation, we use LO cross sections for the production of a heavy Higgs boson
in the context of the Type II 2HDM of Ref.~\cite{Craig:2016ygr}.
The mass of the new particle is varied in the range [350, 550]\GeV,
where the lower mass boundary is chosen in such a way as to allow the decay of
the (pseudo)scalar into on-shell top quarks.

\section{ The CMS detector and event reconstruction}
\label{sec:cms}

The central feature of the CMS detector is a superconducting solenoid of 6\unit{m} internal diameter, providing a magnetic field of 3.8\unit{T}. Within the solenoid volume are a silicon pixel and strip tracker, a lead tungstate crystal electromagnetic calorimeter (ECAL), and a brass and scintillator hadron calorimeter (HCAL), each composed of a barrel and two endcap sections. Forward calorimeters extend the pseudorapidity ($\eta$) coverage provided by the barrel and endcap detectors. Muons are measured in gas-ionization detectors embedded in the steel flux-return yoke outside the solenoid.
A more detailed description of the CMS detector, together with a definition of the coordinate system used and the relevant kinematic variables, can be found in Ref.~\cite{Chatrchyan:2008zzk}.

Events of interest are selected using a two-tiered trigger system~\cite{Khachatryan:2016bia}. The first level (L1), composed of custom hardware processors, uses information from the calorimeters and muon detectors to select events at a rate of around 100\unit{kHz} within a time interval of less than 4\mus. The second level, known as the high-level trigger (HLT), consists of a farm of processors running a version of the full event reconstruction software optimized for fast processing, and reduces the event rate to less than 1\unit{kHz} before data storage.

Events are processed using the particle-flow (PF) algorithm~\cite{PFT-09-001,PFT-10-001}, which reconstructs and identifies each individual particle with an optimized combination of information from the various elements of the CMS detector. The energy of photons is directly obtained from the ECAL measurement. The energy of electrons is determined from a combination of the electron momentum at the primary interaction vertex as determined by the tracker, the energy of the corresponding ECAL cluster, and the energy sum of all bremsstrahlung photons spatially compatible with the electron track~\cite{Khachatryan:2015hwa}. The energy of muons is obtained from the curvature of the corresponding track, combining information from the silicon tracker and the muon system~\cite{Chatrchyan:2012xi}. The energy of charged hadrons is determined from a combination of their momentum measured in the tracker and the matching ECAL and HCAL energy deposits, corrected
for the response function of the calorimeters to hadronic showers. Finally, the energy of neutral hadrons is obtained from the corresponding corrected ECAL and HCAL energy.

Hadronic jets are clustered from neutral PF candidates and charged PF candidates associated with the primary vertex, using the anti-\kt algorithm~\cite{Cacciari:2008gp, Cacciari:2011ma} with a distance parameter $R = \sqrt{\smash[b]{ (\Delta\eta)^2 + (\Delta\phi)^2} }$ of 0.4.
Jet momentum is determined as the vectorial sum of all PF candidate momenta in the jet. An offset correction is applied to jet energies to take into account the contribution from additional proton-proton interactions (pileup) within the same or nearby bunch crossings. Jet energy corrections are derived from simulation, and are improved with in situ measurements of the energy balance in dijet and photon+jet events~\cite{Chatrchyan:2011ds,Khachatryan:2016kdb}. Additional selection criteria are applied to each event to remove spurious jet-like features originating from isolated noise patterns in certain HCAL regions.
Jets originating from \PQb quarks are identified (b tagged) using the medium working point of the combined secondary vertex algorithm CSVv2~\cite{CMS-PAS-BTV-15-001}.
The missing transverse momentum vector \ptvecmiss is defined as the projection on the plane perpendicular to the beams of the negative vector sum of the momenta of all reconstructed PF candidates in an event~\cite{CMS-PAS-JME-16-004}. Its magnitude is referred to as \ETmiss. The sum of the transverse momenta of all jets in an event is referred to as \HT.

\section{Event selection and search strategy}
\label{sec:selection}

The event selection and the definition of the signal regions (SRs) follow closely the analysis strategy established in Ref.~\cite{SUS-15-008}.
With respect to the previous search, the general strategy has remained unchanged.
We target, in a generic way, new physics signatures that result in SS dileptons,
hadronic activity, and \MET, by subdividing the event sample into several SRs sensitive to a variety of new physics models.
The number of SRs was increased to take advantage of the larger integrated luminosity.
Table~\ref{tab:objetselection} summarizes the basic kinematic requirements for jets and leptons
(further details, including the lepton identification and isolation requirements, can be found in Ref.~\cite{SUS-15-008}).

\begin{table}[!hbtp]
\centering
\topcaption{Kinematic requirements for leptons and jets. Note that the \pt thresholds to count jets and b-tagged jets are different. }
\label{tab:objetselection}
\begin{tabular}{c|cc}
\hline
Object        & $\pt$ ($\GeV$)           & $|\eta|$ \\
\hline \hline
Electrons     & $>$15 & $< 2.5 $ \\
Muons         & $>$10 & $< 2.4 $ \\
Jets          & $>$40 & $< 2.4 $ \\
b-tagged jets & $>$25 & $< 2.4 $ \\
\hline
\end{tabular}

\end{table}

Events are selected using triggers based on two sets of HLT algorithms, one simply requiring two leptons, and one additionally requiring $\HT > 300\GeV$.
The \HT requirement allows for the lepton isolation requirement to be removed and for the lepton \pt thresholds to be set to 8\GeV
for both leptons, while in the pure dilepton trigger the leading and subleading leptons are required to have $\pt > 23~(17)\GeV$
and $\pt > 12~(8)\GeV$, respectively, for electrons (muons).
Based on these trigger requirements, leptons are classified as high ($\pt > 25\GeV$) and low ($10 < \pt < 25\GeV$) momentum,
and three analysis regions are defined: high-high (HH), high-low (HL), and low-low (LL).

The baseline selection used in this analysis requires at least one SS lepton pair with an invariant mass above 8\GeV,
at least two jets, and  $\MET > 50\GeV$.
To reduce Drell--Yan backgrounds, events are rejected if an additional loose lepton forms an opposite-sign
same-flavor pair with one of the two SS leptons, with an invariant mass less than 12\GeV or between 76 and 106\GeV.
Events passing the baseline selection are then divided into SRs to separate the different background processes and
to maximize the sensitivity to signatures with different jet multiplicity (\Njets), flavor (\Nbjets), visible and invisible energy (\HT and \MET),
and lepton momentum spectra (the HH/HL/LL categories mentioned previously).
The \MTmin variable is defined as the smallest of the transverse masses constructed between \ptvecmiss and each of the leptons.
This variable features a cutoff near the \PW\ boson mass for processes with only one prompt lepton, so it is used to create SRs where
the nonprompt lepton background is negligible.
To further improve sensitivity, several regions are split according to the charge of the leptons ($++$ or $--$),
taking advantage of the charge asymmetry of SM backgrounds, such as $\ttbar \PW$~or \PW\PZ, with a single \PW\ boson produced in $\Pp\Pp$ collisions.
Only signal regions dominated by such backgrounds and with a sufficient predicted yield are split by charge.
In the HH and HL categories, events in the tail regions $\HT > 1125\GeV$ or $\MET > 300\GeV$ are inclusive in
 \Njets, \Nbjets, and \MTmin in order to ensure a reasonable yield of events in these SRs.
The exclusive SRs resulting from this classification are defined in Tables~\ref{tab:SRDefHH}--\ref{tab:SRDefLL}.

The lepton reconstruction and identification efficiency is in the range of 45--70\% (70--90\%)  for electrons (muons)
with $\pt>25\GeV$, increasing as a function of \pt and converging to the maximum value for $\pt>60\GeV$.
In the low-momentum regime, $15<\pt<25\GeV$ for electrons and $10<\pt<25\GeV$ for muons,
the efficiencies are 40\% for electrons and 55\% for muons.
The lepton trigger efficiency for electrons is in the range of 90-98\%, converging to the maximum
value for $\pt>30\GeV$, and around 92\% for muons.
The chosen b tagging working point results in approximately a 70\% efficiency for tagging a \PQb quark jet and a $<$1\%
mistagging rate for light-flavor jets in \ttbar events~\cite{CMS-PAS-BTV-15-001}.
The efficiencies of the \HT and \ETmiss requirements are mostly determined by the jet energy and \MET resolutions,
which are discussed in Refs.~\cite{Chatrchyan:2011ds,Khachatryan:2016kdb,JME-13-003}.

\begin{table*}
  \centering
      \topcaption{\label{tab:SRDefHH}  Signal region definitions for the HH selection. Regions split by charge are indicated with ($++$) and ($--$). }
\resizebox*{1\textwidth}{!}{
\begin{tabular}{|c|c|c|c|c|c|c|c|c|}
\hline
$\Nbjets$                         & $\MTmin$ (\GeVns{})            & $\ETmiss$ (\GeVns{})            & $\Njets$            & $\HT < 300\GeV$        & $\HT\in[300, 1125]\GeV$         & $\HT\in[1125, 1300]\GeV$ & $\HT\in[1300, 1600]\GeV$ & $\HT > 1600\GeV$  \\ \hline \hline
    \multirow{8}{*}{0}            & \multirow{4}{*}{$<$120}    & \multirow{2}{*}{ $50-200$}  & 2-4                 & SR1                    & SR2                             & \multirow{30}{*}{\begin{tabular}{l} SR46 ($++$) / \\ SR47 ($--$) \end{tabular}}  & \multirow{30}{*}{\begin{tabular}{l} SR48 ($++$) / \\SR49 ($--$) \end{tabular}}  & \multirow{30}{*}{\begin{tabular}{l} SR50 ($++$) / \\ SR51 ($--$) \end{tabular}} \\ \cline{4-6}
                                  &                            &                             & $\geq$5                  & \multirow{7}{*}{SR3}   & SR4                             &                          &                          & \\ \cline{3-4} \cline{6-6}
                                  &                            & \multirow{2}{*}{$200-300$}  & 2-4                 &                        & SR5 ($++$) / SR6 ($--$)                            &                          &                          & \\ \cline{4-4} \cline{6-6}
                                  &                            &                             & $\geq$5                  &                        & SR7                             &                          &                          & \\ \cline{2-4} \cline{6-6}
                                  & \multirow{4}{*}{$>$120}    & \multirow{2}{*}{ $50-200$}  & 2-4                 &                        & SR8 ($++$) / SR9 ($--$)                            &                          &                          & \\ \cline{4-4} \cline{6-6}
                                  &                            &                             & $\geq$5                  &                        & \multirow{3}{*}{SR10}           &                          &                          & \\ \cline{3-4}
                                  &                            & \multirow{2}{*}{$200-300$}  & 2-4                 &                        &                                 &                          &                          & \\ \cline{4-4}
                                  &                            &                             & $\geq$5                  &                        &                                 &                          &                          & \\ \cline{1-6}
    \multirow{8}{*}{1}            & \multirow{4}{*}{$<$120}    & \multirow{2}{*}{ $50-200$}  & 2-4                 & SR11                   & SR12                            &                          &                          & \\ \cline{4-6}
                                  &                            &                             & $\geq$5                  & \multirow{7}{*}{\begin{tabular}{l} SR13 ($++$) / \\ SR14 ($--$) \end{tabular}} & SR15 ($++$) / SR16 ($--$)                           &                          &                          & \\ \cline{3-4} \cline{6-6}
                                  &                            & \multirow{2}{*}{ $200-300$} & 2-4                 &                        & SR17 ($++$) / SR18 ($--$)                           &                          &                          & \\ \cline{4-4} \cline{6-6}
                                  &                            &                             & $\geq$5                  &                        & SR19                            &                          &                          & \\ \cline{2-4} \cline{6-6}
                                  & \multirow{4}{*}{$>$120}    & \multirow{2}{*}{ $50-200$}  & 2-4                 &                        & SR20 ($++$) / SR21 ($--$)                           &                          &                          & \\ \cline{4-4} \cline{6-6}
                                  &                            &                             & $\geq$5                  &                        & \multirow{3}{*}{SR22}           &                          &                          & \\ \cline{3-4}
                                  &                            & \multirow{2}{*}{ $200-300$} & 2-4                 &                        &                                 &                          &                          & \\ \cline{4-4}
                                  &                            &                             & $\geq$5                  &                        &                                 &                          &                          & \\ \cline{1-6}
    \multirow{8}{*}{2}            & \multirow{4}{*}{$<$120}    & \multirow{2}{*}{ $50-200$}  & 2-4                 & SR23                   & SR24                            &                          &                          & \\ \cline{4-6}
                                  &                            &                             & $\geq$5                  & \multirow{7}{*}{\begin{tabular}{l} SR25 ($++$) / \\ SR26 ($--$) \end{tabular}} & SR27 ($++$) / SR28 ($--$)                           &                          &                          & \\ \cline{3-4} \cline{6-6}
                                  &                            & \multirow{2}{*}{ $200-300$} & 2-4                 &                        & SR29 ($++$) / SR30 ($--$)                           &                          &                          & \\ \cline{4-4}  \cline{6-6}
                                  &                            &                             & $\geq$5                  &                        & SR31                            &                          &                          & \\ \cline{2-4} \cline{6-6}
                                  & \multirow{4}{*}{$>$120}    & \multirow{2}{*}{ $50-200$}  & 2-4                 &                        & SR32 ($++$) / SR33 ($--$)                           &                          &                          & \\ \cline{4-4} \cline{6-6}
                                  &                            &                             & $\geq$5                  &                        & \multirow{3}{*}{SR34}           &                          &                          & \\ \cline{3-4}
                                  &                            & \multirow{2}{*}{ $200-300$} & 2-4                 &                        &                                 &                          &                          & \\ \cline{4-4}
                                  &                            &                             & $\geq$5                  &                        &                                 &                          &                          & \\ \cline{1-6}
    \multirow{4}{*}{$\geq$3}           & \multirow{2}{*}{$<$120}    & $50-200$                    & \multirow{2}{*}{$\geq$2} & \multirow{2}{*}{\begin{tabular}{l} SR35 ($++$) / \\ SR36 ($--$) \end{tabular}} & SR37 ($++$) / SR38 ($--$)                           &                          &                          & \\ \cline{3-3} \cline{6-6}
                                  &                            & $200-300$                   &                     &                        & SR39                            &                          &                          & \\ \cline{2-6}
                                  & \multirow{2}{*}{$>$120}    & \multirow{2}{*}{$50-300$}     & \multirow{2}{*}{$\geq$2} & \multirow{2}{*}{SR40}  & \multirow{2}{*}{SR41}           &                          &                          & \\                                     &  &  &  &  &  &  &  & \\
    \hline \multirow{2}{*}{inclusive} & \multirow{2}{*}{inclusive} & $300-500$                   & \multirow{2}{*}{$\geq$2} & \NA      & \multicolumn{4}{c|}{SR42 ($++$) / SR43 ($--$)}  \\
    \cline{3-3} \cline{6-9}           &                            & $>$500                      &                     & \NA      & \multicolumn{4}{c|}{SR44 ($++$) / SR45 ($--$)}  \\
\hline
\end{tabular}
}
\end{table*}

\begin{table*}[h!]
  \centering
    \topcaption{\label{tab:SRDefHL} Signal region definitions for the HL selection. Regions split by charge are indicated with ($++$) and ($--$). }
\resizebox*{1\textwidth}{!}{
\begin{tabular}{|c|c|c|c|c|c|c|c|}
\hline
$\Nbjets$ & $\MTmin$ (\GeVns{})& $\ETmiss$ (\GeVns{})& $\Njets$ & $\HT < 300\GeV$ &  $\HT\in[300, 1125]\GeV$ & $\HT\in[1125, 1300]\GeV$ & $\HT > 1300\GeV$  \\
\hline \hline
    \multirow{4}{*}{0}         & \multirow{4}{*}{$<$120}    & \multirow{2}{*}{$50-200$}   & 2-4                 & SR1                   & SR2                            & \multirow{17}{*}{\begin{tabular}{l} SR38 ($++$) / \\ SR39 ($--$) \end{tabular}}   & \multirow{17}{*}{\begin{tabular}{l} SR40 ($++$) / \\ SR41 ($--$) \end{tabular}} \\
\cline{4-6}
                           &                            &                             & $\geq$5                  & \multirow{3}{*}{SR3}  & SR4                            &                          & \\
\cline{3-4} \cline{6-6}
                           &                            & \multirow{2}{*}{$200-300$}  & 2-4                 &                       & SR5 ($++$) / SR6 ($--$)                            &                          & \\
\cline{4-4} \cline{6-6}
                           &                            &                             & $\geq$5                  &                       & SR7                            &                          & \\
\cline{1-6}
\multirow{4}{*}{1}         & \multirow{4}{*}{$<$120}    & \multirow{2}{*}{ $50-200$}  & 2-4                 & SR8                   & SR9                            &                          & \\
\cline{4-6}
                           &                            &                             & $\geq$5                  & \multirow{3}{*}{\begin{tabular}{l} SR10 ($++$) / \\ SR11 ($--$) \end{tabular}}  & SR12 ($++$) / SR13 ($--$)                           &                          & \\
\cline{3-4} \cline{6-6}
                           &                            & \multirow{2}{*}{ $200-300$} & 2-4                 &                       & SR14 ($++$) / SR15 ($--$)                           &                          & \\
\cline{4-4} \cline{6-6}
                           &                            &                             & $\geq$5                  &                       & SR16 ($++$) / SR17 ($--$)                           &                          & \\
\cline{1-6}
\multirow{4}{*}{2}         & \multirow{4}{*}{$<$120}    & \multirow{2}{*}{ $50-200$}  & 2-4                 & SR18                  & SR19                           &                          & \\
\cline{4-6}
                           &                            &                             & $\geq$5                  & \multirow{3}{*}{\begin{tabular}{l} SR20 ($++$) / \\ SR21 ($--$) \end{tabular}} & SR22 ($++$) / SR23 ($--$)                           &                          & \\
\cline{3-4} \cline{6-6}
                           &                            & \multirow{2}{*}{ $200-300$} & 2-4                 &                       & SR24 ($++$) / SR25 ($--$)                           &                          & \\
\cline{4-4}  \cline{6-6}
                           &                            &                             & $\geq$5                  &                       & SR26                           &                          & \\
\cline{1-6}
    \multirow{2}{*}{$\geq$3}        & \multirow{2}{*}{$<$120}    & $50-200$                    & \multirow{2}{*}{$\geq$2} & \multirow{2}{*}{\begin{tabular}{l} SR27 ($++$) / \\ SR28 ($--$) \end{tabular}} & SR29 ($++$) / SR30 ($--$)                           &                          & \\
\cline{3-3} \cline{6-6}
                           &                            & $200-300$                   &                     &                       & SR31                           &                          & \\
\cline{1-6}
inclusive                  & $>$120                     & $50-300$                    & $\geq$2                  & SR32                  & SR33                           &                          & \\
\hline
    \multirow{2}{*}{inclusive} & \multirow{2}{*}{inclusive} & $300-500$                   & \multirow{2}{*}{$\geq$2} & \NA    & \multicolumn{3}{c|}{SR34 ($++$) / SR35 ($--$)}  \\
\cline{3-3} \cline{6-8}
                           &                            & $>$500                      &                     & \NA    & \multicolumn{3}{c|}{SR36 ($++$) / SR37 ($--$)}  \\
\hline
\end{tabular}
}
\end{table*}

\begin{table*}
  \centering
    \topcaption{\label{tab:SRDefLL} Signal region definitions for the LL selection. All SRs in this category require $\Njets \geq 2$.}
\begin{tabular}{|c|c|c|c|c|}
\hline
$\Nbjets$ &  $\MTmin$ (\GeVns{})& $\HT$ (\GeVns{})& $\ETmiss\in[50, 200]\GeV$ &  $\ETmiss> 200\GeV$ \\
\hline\hline
0 & \multirow{4}{*}{$<$120} &\multirow{5}{*}{$>$300} & SR1 & SR2 \\
\cline{1-1} \cline{4-5}
1 & & & SR3 & SR4 \\
\cline{1-1} \cline{4-5}
2 & & & SR5 & SR6 \\
\cline{1-1} \cline{4-5}
$\geq$3 & & & \multicolumn{2}{c|}{SR7} \\
\cline{1-2} \cline{4-5}
Inclusive & $>$120 & & \multicolumn{2}{c|}{SR8} \\
\hline
\end{tabular}
\end{table*}

\section{Backgrounds}
\label{sec:backgrounds}
Standard model background contributions arise from three sources:
processes with prompt SS dileptons,
mostly relevant in regions with high \MET or \HT; events with a nonprompt lepton, dominating the overall final state;
and opposite-sign dilepton events with a charge-misidentified lepton, the smallest contribution.
In this paper we use the shorthand ``nonprompt leptons'' to refer to electrons or muons from the decays of heavy- or light-flavor hadrons,
hadrons misidentified as leptons, or electrons from conversions of photons in jets.

Several categories of SM processes that result in the production of electroweak bosons can give rise to an SS dilepton final state.
These include production of multiple bosons in the same
event (prompt photons, \PW, \PZ, and Higgs bosons),  as well as single-boson production in association with top quarks.
Among these SM processes, the dominant ones are \PW\PZ,  $\ttbar \PW$, and $\ttbar \PZ$ production, followed by the $\PW^{\pm}\PW^{\pm}$ process.
The remaining SM processes are grouped into two categories,
``Rare'' (including $\PZ\PZ$, \PW\PW\PZ, $\PW\PZ\PZ$, $\PZ\PZ\PZ$, $\PQt\PW\PZ$, $\PQt\PZ\PQq$, as well as $\ttbar\ttbar$ and double parton scattering)
and ``X+\Pgg'' (including $\PW\Pgg$, $\PZ\Pgg$, $\ttbar \Pgg$, and $\PQt\Pgg$).
The expected yields from these SM backgrounds are estimated from simulation, accounting for both the theoretical and experimental uncertainties discussed in Section~\ref{sec:systematics}.

For the \PW\PZ and $\ttbar \PZ$ backgrounds, a three-lepton (3L) control region in data is used to scale the simulation, based on a template
fit to the distribution of the number of b jets.
The 3L control region requires at least two jets, $\MET > 30\GeV$, and three leptons, two of which must form an opposite-sign same-flavor
pair with an invariant mass within $15\GeV$ of the \PZ boson mass.
In the fit to data, the normalization and shapes of all the components are allowed to vary
according to experimental and theoretical uncertainties.
The scale factors obtained from the fit in the phase space of the 3L control region
are $1.26 \pm 0.09$ for the \PW\PZ process, and $1.14 \pm 0.30$ for the $\ttbar \PZ$ process.

The nonprompt lepton background, which is largest for regions with low \MTmin and low \HT,
is estimated by the ``tight-to-loose'' method, which was employed in several previous versions of the analysis~\cite{SUS-10-004,SUS-11-020,SUS-11-010,SUS-12-017,SUS-13-013},
and significantly improved in the latest version~\cite{SUS-15-008}
to account for the kinematics and flavor of the parent parton of the nonprompt lepton.
The tight-to-loose method uses two control regions, the measurement region and the application region.
The measurement region consists of a sample of single-lepton events enriched in nonprompt leptons by requirements on 
\MET and transverse mass that suppress the $\PW \to \ell \nu$ contribution.  This sample is used to extract the probability
for a nonprompt lepton that satisfies the loose selection to also satisfy the tight selection.  This probability ($\epsilon_{\rm TL}$) is 
calculated as a function of  lepton $\pt^\text{corr}$ (defined below) and $\eta$, 
separately for electrons and muons, and separately for lepton triggers with and without an isolation
requirement.  The application region is a SS dilepton region where both of the leptons satisfy the loose selection but at least one 
of them fails the tight selection.  This region is subsequently divided into a set of subregions with the exact same kinematic requirements
as those in the SRs.  Events in the subregions are weighted by a factor $\epsilon_{\rm TL} / (1-\epsilon_{\rm TL})$ for each lepton in the event failing the 
tight requirement. The nonprompt background in each SR is then estimated as the sum of the event weights in the corresponding subregion.
The $\pt^\text{corr}$ parametrization, where $\pt^\text{corr}$ is defined as the lepton \pt plus the energy in the
isolation cone exceeding the isolation threshold value, is chosen because of its correlation with the parent parton \pt,
improving the stability of  the $\epsilon_\mathrm{TL}$ values with respect to the sample kinematics.
To improve the stability of the $\epsilon_\mathrm{TL}$ values with respect to the flavor of the parent parton, the loose electron selection is adopted.
This selection increases the number of nonprompt electrons from the fragmentation and decay of light-flavor partons, resulting
in $\epsilon_\mathrm{TL}$ values similar to those from heavy-flavor parent partons.

The prediction from the tight-to-loose method is cross-checked using an alternative method based on the same principle,
similar to that described in Ref.~\cite{ATLAS:2014kca}. In this cross-check, which aims to remove kinematic differences
between measurement and application regions, the measurement region is obtained from SS dilepton events
where one of the leptons fails the impact parameter requirement. With respect to the nominal method, the loose lepton definition
is adapted to reduce the effect of the correlation between isolation and impact parameter.
The predictions of the two methods are found to be consistent within systematic uncertainties.

Charge misidentification of electrons is a small background that can arise from severe brems\-strah\-lung in the tracker material. Simulation-based studies with tight leptons indicate that the muon charge misidentification probability is negligible, while for electrons it ranges between $10^{-5}$ and $10^{-3}$. The charge misidentification background is estimated from data using an opposite-sign control region for each SS SR, scaling the control region yield by the charge misidentification probability measured in simulation. A low-\MET control region, with $\Pep\Pem$ pairs in the \PZ boson mass window, is used to cross-check the MC prediction for the misidentification probability, both inclusively and --- where the number of events in data allows it --- as a function of electron \pt and $\eta$.

\section{Systematic uncertainties}
\label{sec:systematics}

Several sources of systematic uncertainty affect the predicted yields for signal and background processes,
as summarized in Table~\ref{tab:systSummary}.
Experimental uncertainties are based on measurements in data of the trigger efficiency, the lepton identification efficiency,
the b tagging efficiency~\cite{CMS-PAS-BTV-15-001},
the jet energy scale, and the integrated luminosity~\cite{CMS-PAS-LUM-17-001},
as well as on the inelastic cross section value affecting the pileup rate.
Theoretical uncertainties related to unknown higher-order effects are
estimated by varying simultaneously the factorization
and renormalization scales by a factor of two,
while uncertainties in the PDFs are obtained using replicas of the NNPDF3.0 set~\cite{Ball:2014uwa}.

Experimental and theoretical uncertainties affect both the overall yield (normalization) and the relative population
(shape) across SRs, and they are taken into account for all signal samples as well as for
the samples used to estimate the main prompt SS dilepton backgrounds:
\PW\PZ, $\ttbar\PW$,
$\ttbar\PZ$, $\PW^{\pm} \PW^{\pm}$.
For the \PW\PZ and $\ttbar\PZ$ backgrounds,  the control region fit results are used for the normalization,
so these uncertainties are only taken into account for the shape of the backgrounds.
For the smallest background samples, Rare and X+\Pgg, a 50\% uncertainty is assigned
in place of the scale and PDF variations.

The normalization and the shapes of the nonprompt lepton and charge misidentification backgrounds are estimated from control regions in data.
In addition to the statistical uncertainties from the control region yields, dedicated systematic uncertainties are associated with the methods used in this estimate.
For the nonprompt lepton background, a 30\% uncertainty (increased to 60\% for electrons with $\pt > 50\GeV$) accounts for the
performance of the method in simulation and for the differences in the two alternative methods described in
Section~\ref{sec:backgrounds}.
In addition, the uncertainty in the prompt lepton yield in the measurement region,
relevant when estimating $\epsilon_\mathrm{TL}$ for high-\pt leptons, results in a 1--30\% effect on the estimate.
For the charge misidentification background, a 20\% uncertainty is assigned to account for possible mismodeling
of the charge misidentification rate in simulation.

\begin{table}
\centering
    \topcaption{
   Summary of the sources of uncertainty and their effect on the yields of different processes in the SRs.
    The first two groups list experimental and theoretical uncertainties assigned to processes estimated using simulation,
    while the last group lists uncertainties assigned to processes whose yield is estimated from data.
    The uncertainties in the first group also apply to signal samples.
    Reported values are representative for the most relevant signal regions.
    }
    \label{tab:systSummary}
\begin{tabular}{l|c}
\hline
Source          & Typical uncertainty (\%) \\
\hline\hline
Integrated luminosity & 2.5 \\
Lepton selection & $4-10$ \\
Trigger efficiency & $2-7$ \\
Pileup & $0-6$ \\
Jet energy scale & $1-15$ \\
$\PQb$ tagging & $1-15$ \\
Simulated sample size & $1-10$ \\
\hline
Scale and PDF variations & $10-20$ \\
\hline
\PW\PZ (normalization) & 12 \\
$\ttbar \PZ$ (normalization) & 30 \\
Nonprompt leptons & $30-60$ \\
Charge misidentification & 20 \\
\hline
\end{tabular}

\end{table}

\section{Results and interpretation}
\label{sec:results}

\begin{figure*}[htbp]
\centering
\includegraphics[width=0.35\textwidth]{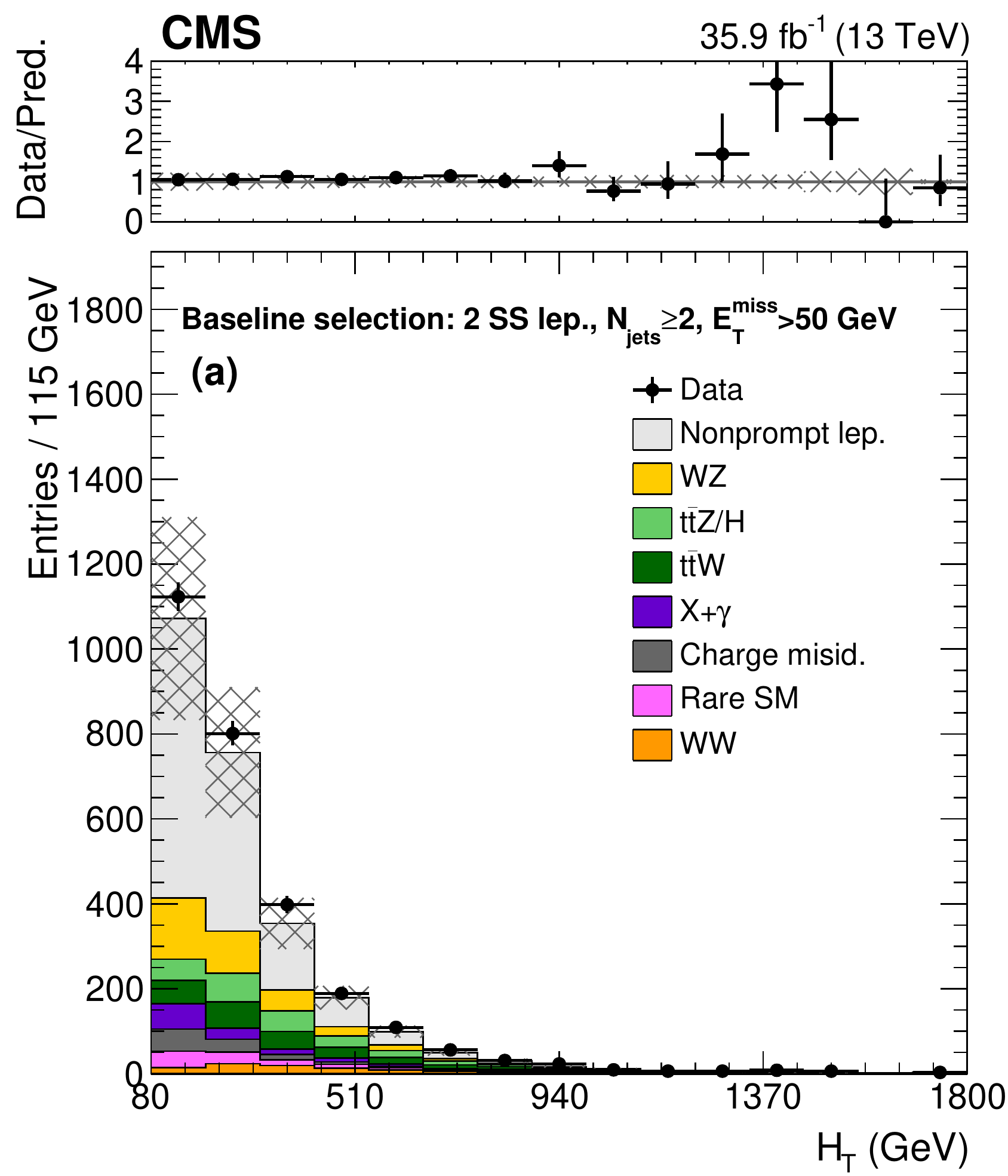}
\\
\includegraphics[width=0.35\textwidth]{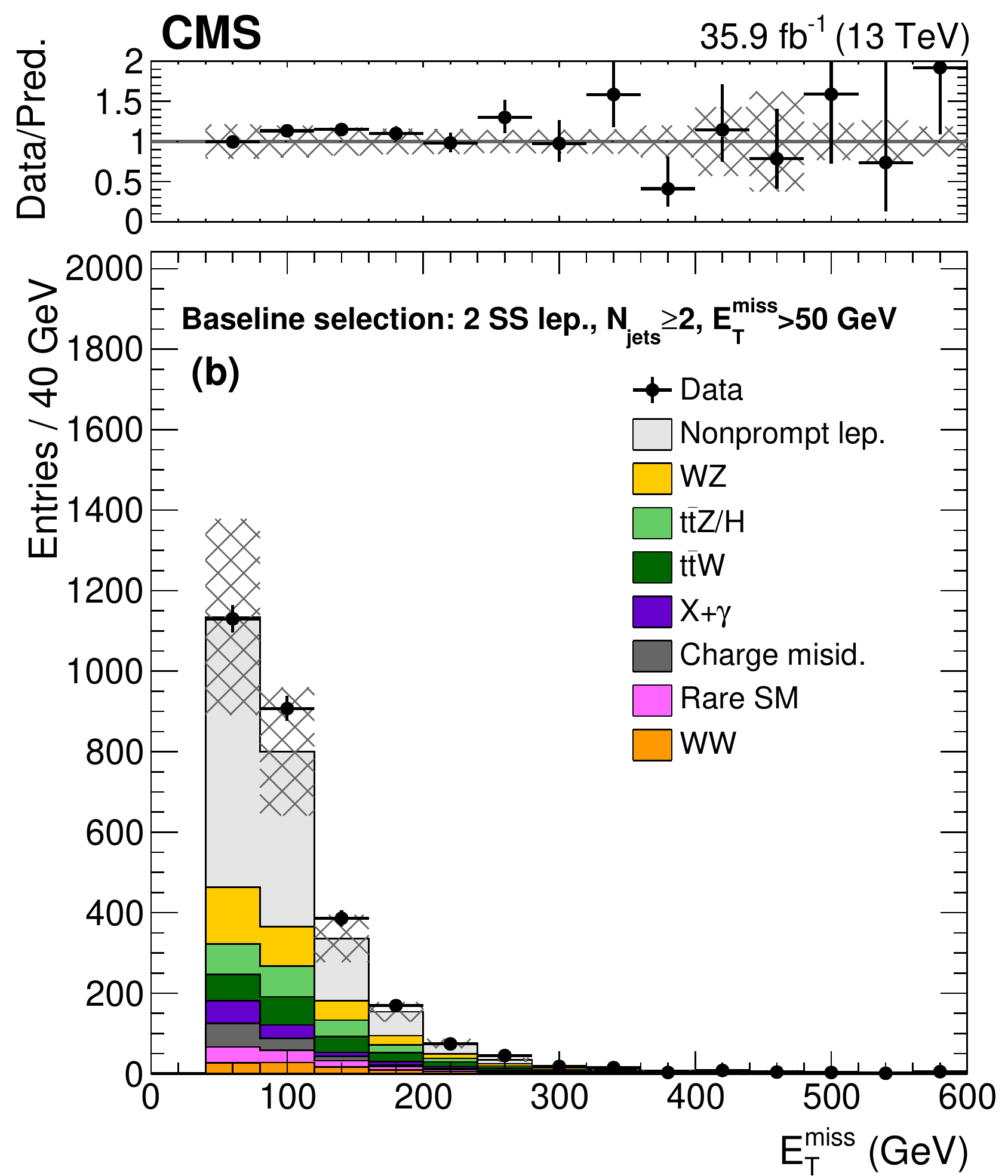}
\includegraphics[width=0.35\textwidth]{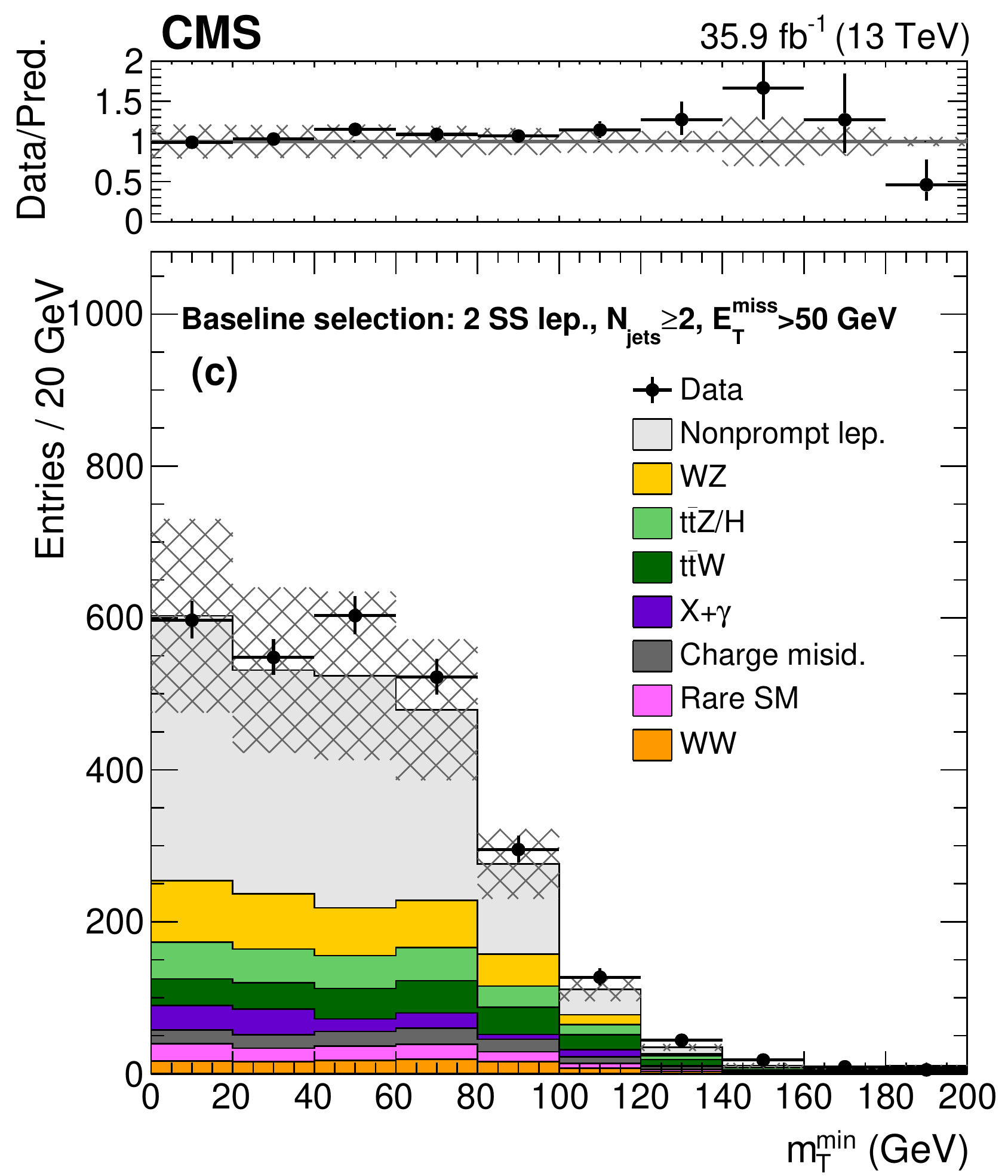}
\\
\includegraphics[width=0.35\textwidth]{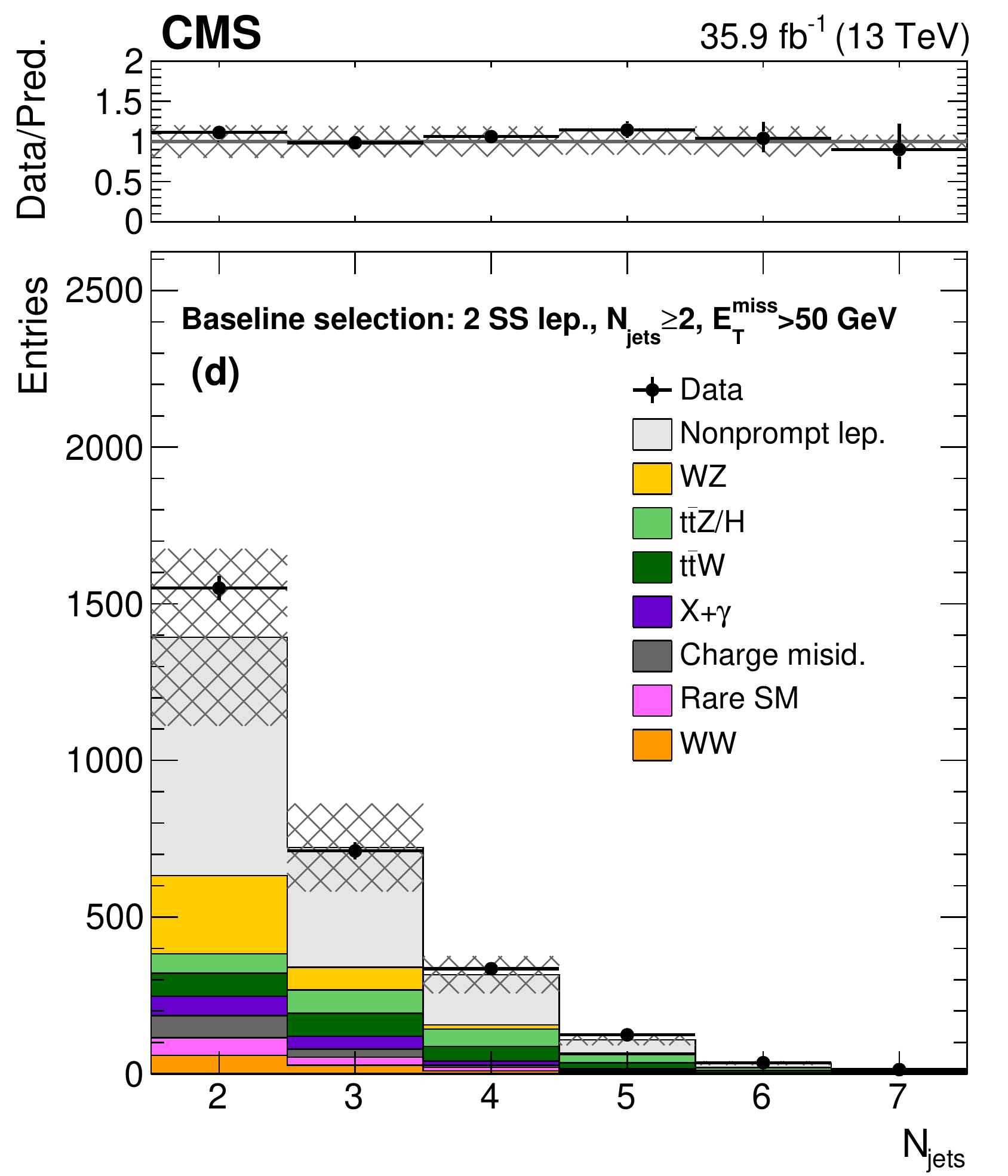}
\includegraphics[width=0.35\textwidth]{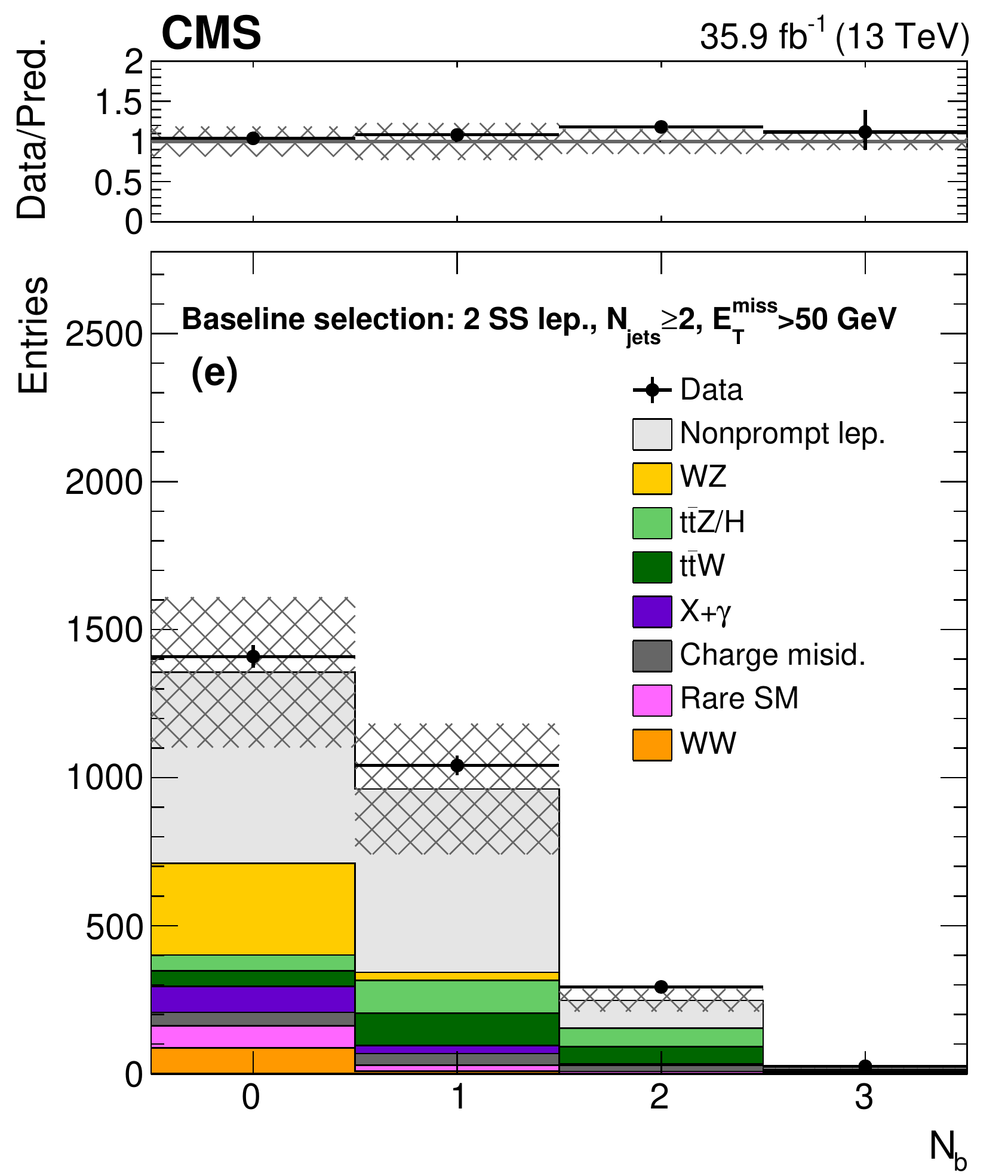}
\caption{  Distributions of the main analysis variables:
\HT~(a), \ETmiss~(b), \MTmin~(c), \Njets~(d), and \Nbjets~(e),
after the baseline selection requiring a pair of SS leptons, two jets, and $\MET > 50\GeV$.
The last bin includes the overflow events and the hatched area represents the total uncertainty in the background prediction.
The upper panels show the ratio of the observed event yield to the background prediction. }
\label{fig:kinem}
\end{figure*}

\begin{figure*}[!hbtp]
\centering
\includegraphics[width=0.48\textwidth]{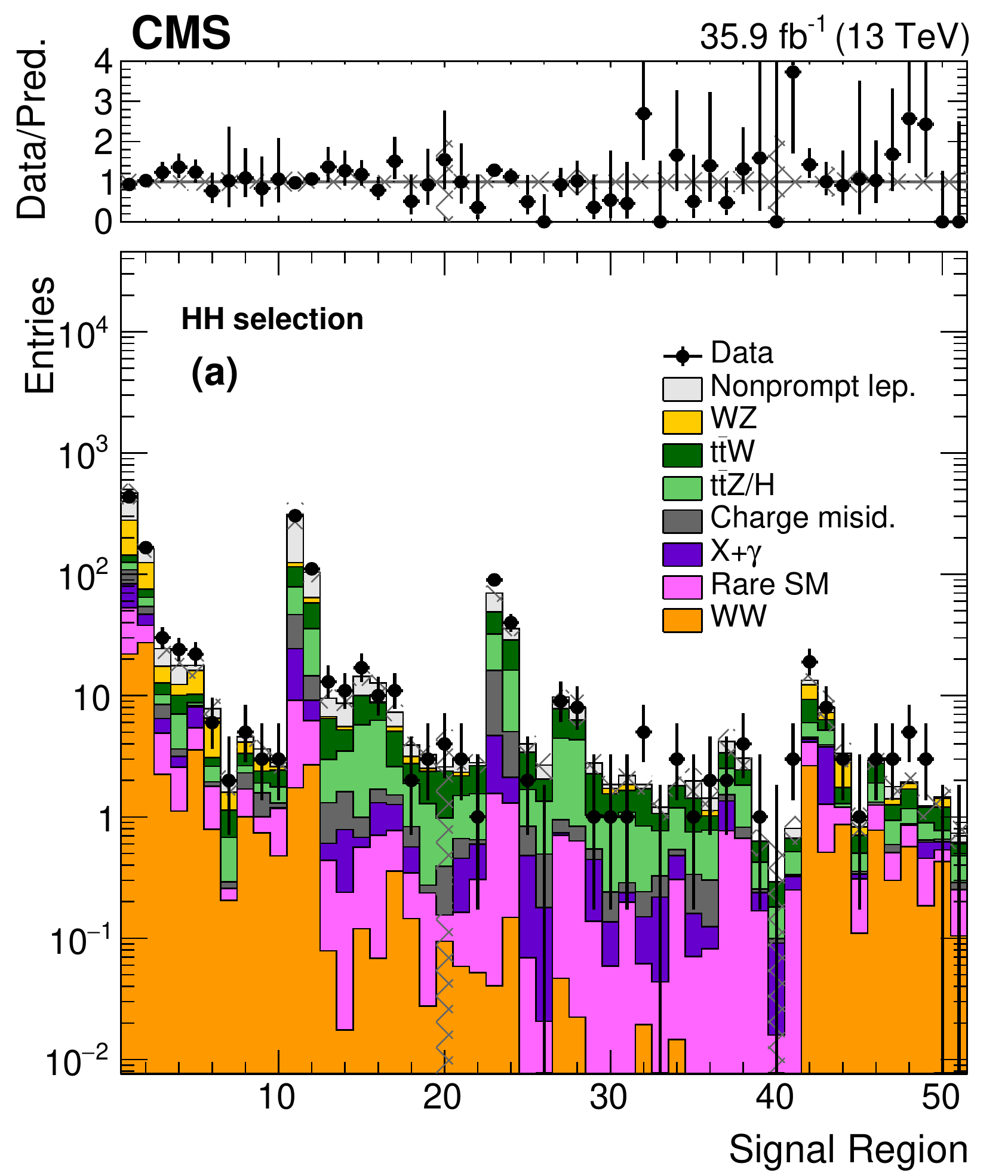}
\includegraphics[width=0.48\textwidth]{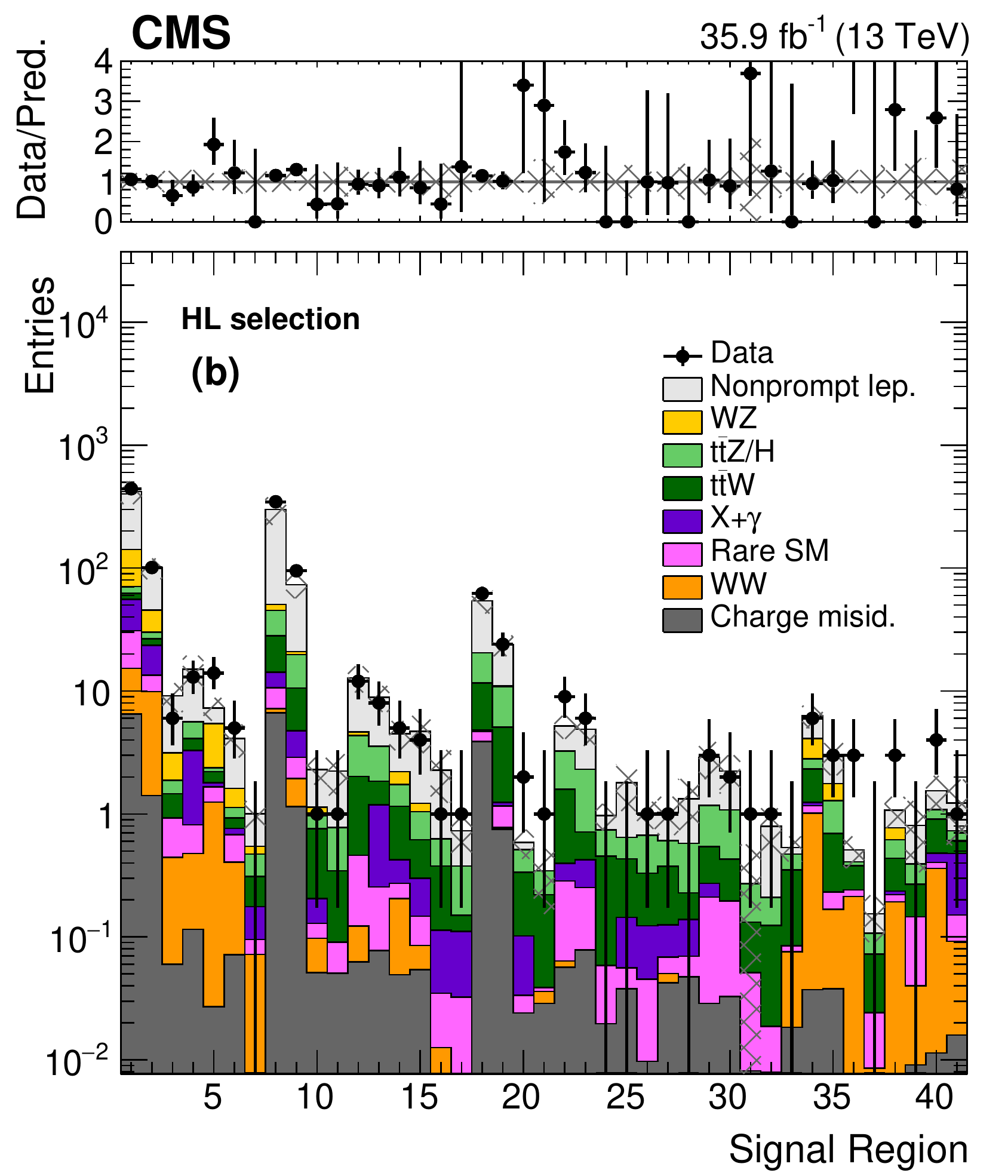}
\includegraphics[width=0.48\textwidth]{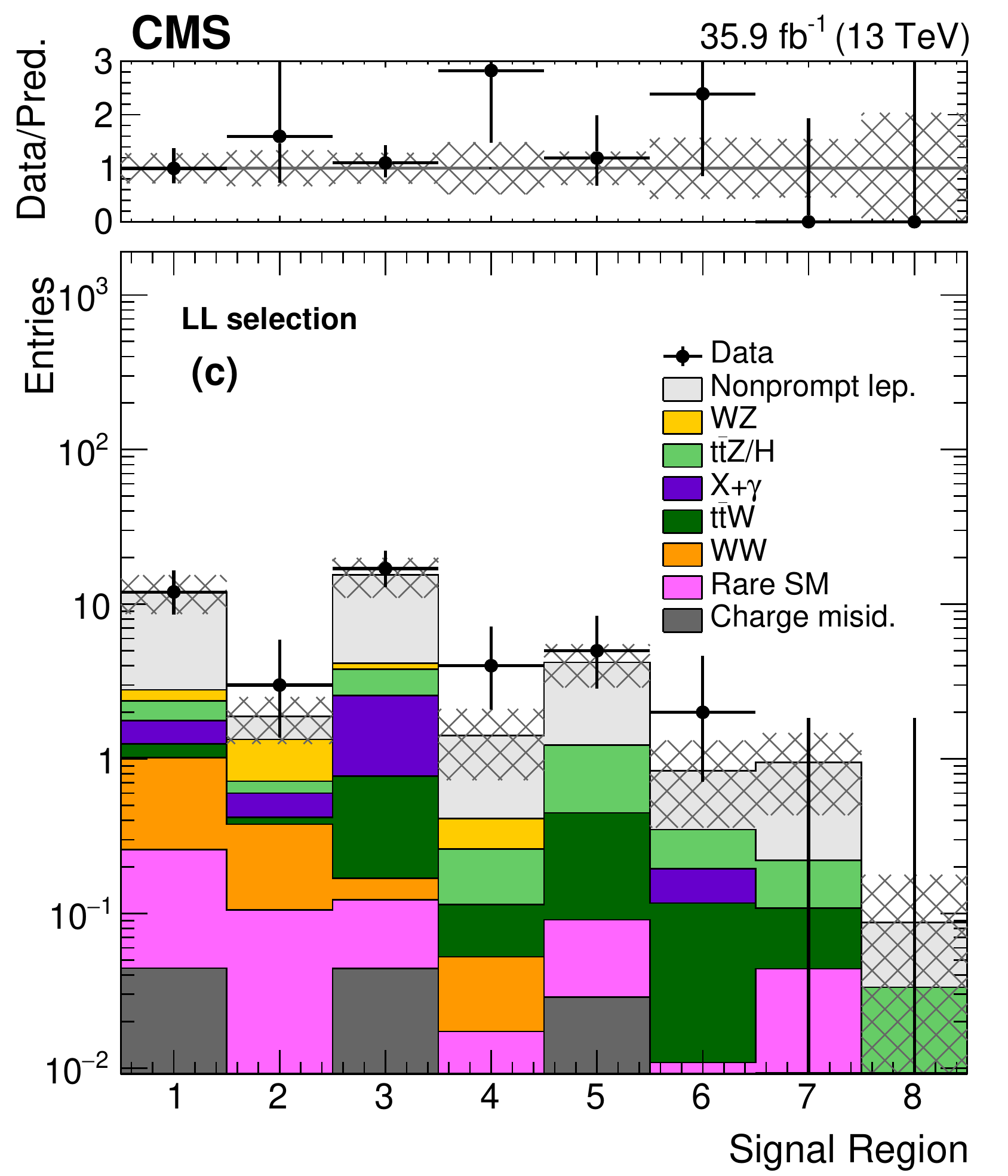}
\caption{ Event yields in the HH~(a), HL~(b), and LL~(c) signal regions.
The hatched area represents the total uncertainty in the background prediction.
The upper panels show the ratio of the observed event yield to the background prediction. }
\label{fig:SR}
\end{figure*}

A comparison between observed yields and the SM background prediction is shown in Fig.~\ref{fig:kinem}
for the kinematic variables used to define the analysis SRs: \HT, \ETmiss, \MTmin, \Njets, and \Nbjets.
The distributions are shown after the baseline selection defined in Section~\ref{sec:selection}.
The full results of the search in each SR
are shown in Fig.~\ref{fig:SR} and Table~\ref{tab:yieldsSR}.
The SM predictions are generally consistent with the data.
The largest deviations are seen in HL SR 36 and 38, with a local significance,
taking these regions individually or combining them with other regions
adjacent in phase space, that does not exceed 2 standard deviations.

These results are used to probe
the signal models discussed in Section~\ref{sec:samples}:
simplified SUSY models, (pseudo)scalar boson production, four top
quark production, and SS top quark production.  We also interpret the
results as model-independent limits as a function of \HT and \MET.
With the exception of the new (pseudo)scalar boson limits, the results can be compared to the previous version of the analysis~\cite{SUS-15-008},
showing significant improvements due to the increase in the integrated luminosity and the optimization of SR definitions.

To obtain exclusion limits at the 95\% confidence level (CL), the results from
all SRs --- including
signal and background uncertainties and their correlations --- are combined using an
asymptotic formulation of the modified frequentist CL$_\mathrm{s}$
criterion~\cite{Junk:1999kv,Read:2002hq,ATL-PHYS-PUB-2011-011,Cowan:2010js}.
When testing a model,
all new particles not included in the specific model are considered too heavy to take part in the interaction.
To convert cross section limits into mass limits, the signal cross sections specified in Section~\ref{sec:samples} are used.

\begin{table*}[!hbtp]
\footnotesize
\centering
\topcaption{Number of expected background and observed events in different SRs in this analysis.}
\label{tab:yieldsSR}
\begin{tabular}{c|cc|cc|cc}
\hline
     & \multicolumn{2}{c|}{HH regions} & \multicolumn{2}{c|}{HL regions} & \multicolumn{2}{c}{LL regions} \\
\hline
       & Expected SM       & Observed & Expected SM       & Observed & Expected SM        & Observed \\
\hline \hline
SR1  & 468 $\pm$ 98 & 435  & 419 $\pm$ 100 & 442 &  12.0 $\pm$  3.9 &  12 \\
SR2  & 162 $\pm$ 25 & 166  & 100 $\pm$ 20 & 101 &  1.88 $\pm$  0.62 &  3 \\
SR3  & 24.4 $\pm$ 5.4 & 30  & 9.2 $\pm$ 2.4 & 6 &  15.5 $\pm$  4.7 &  17 \\
SR4  & 17.6 $\pm$ 3.0 & 24  & 15.0 $\pm$ 4.5 & 13 &  1.42 $\pm$  0.69 &  4 \\
SR5  & 17.8 $\pm$ 3.9 & 22  & 7.3 $\pm$ 1.5 & 14 &  4.2 $\pm$  1.4 &  5 \\
SR6  & 7.8 $\pm$ 1.5 & 6  & 4.1 $\pm$ 1.2 & 5 &  0.84 $\pm$  0.48 &  2 \\
SR7  & 1.96 $\pm$ 0.47 & 2  & 1.01 $\pm$ 0.28 & 0 &  0.95 $\pm$  0.52 &  0 \\
SR8  & 4.58 $\pm$ 0.81 & 5  & 300 $\pm$ 82 & 346 &  0.09 $\pm$  0.07 &  0 \\
SR9  & 3.63 $\pm$ 0.75 & 3  & 73 $\pm$ 17 & 95 &  &   \\
SR10  & 2.82 $\pm$ 0.56 & 3  & 2.30 $\pm$ 0.61 & 1 &  &   \\
SR11  & 313 $\pm$ 87 & 304  & 2.24 $\pm$ 0.87 & 1 &  &   \\
SR12  & 104 $\pm$ 20 & 111  & 12.8 $\pm$ 3.3 & 12 &  &   \\
SR13  & 9.5 $\pm$ 1.9 & 13  & 8.9 $\pm$ 2.3 & 8 &  &   \\
SR14  & 8.7 $\pm$ 2.0 & 11  & 4.5 $\pm$ 1.3 & 5 &  &   \\
SR15  & 14.4 $\pm$ 2.9 & 17  & 4.7 $\pm$ 1.6 & 4 &  &   \\
SR16  & 12.7 $\pm$ 2.6 & 10  & 2.3 $\pm$ 1.1 & 1 &  &   \\
SR17  & 7.3 $\pm$ 1.2 & 11  & 0.73 $\pm$ 0.29 & 1 &  &   \\
SR18  & 3.92 $\pm$ 0.79 & 2  & 54 $\pm$ 12 & 62 &  &   \\
SR19  & 3.26 $\pm$ 0.74 & 3  & 23.7 $\pm$ 4.9 & 24 &  &   \\
SR20  & 2.6 $\pm$ 2.7 & 4  & 0.59 $\pm$ 0.17 & 2 &  &   \\
SR21  & 3.02 $\pm$ 0.75 & 3  & 0.34 $\pm$ 0.20 & 1 &  &   \\
SR22  & 2.80 $\pm$ 0.57 & 1  & 5.2 $\pm$ 1.2 & 9 &  &   \\
SR23  & 70 $\pm$ 12 & 90  & 4.9 $\pm$ 1.4 & 6 &  &   \\
SR24  & 35.7 $\pm$ 5.9 & 40  & 0.97 $\pm$ 0.27 & 0 &  &   \\
SR25  & 3.99 $\pm$ 0.73 & 2  & 1.79 $\pm$ 0.74 & 0 &  &   \\
SR26  & 2.68 $\pm$ 0.80 & 0  & 1.01 $\pm$ 0.27 & 1 &  &   \\
SR27  & 9.7 $\pm$ 1.8 & 9  & 1.03 $\pm$ 0.44 & 1 &  &   \\
SR28  & 7.9 $\pm$ 2.5 & 8  & 1.33 $\pm$ 0.61 & 0 &  &   \\
SR29  & 2.78 $\pm$ 0.58 & 1  & 2.89 $\pm$ 0.99 & 3 &  &   \\
SR30  & 1.86 $\pm$ 0.38 & 1  & 2.24 $\pm$ 0.79 & 2 &  &   \\
SR31  & 2.20 $\pm$ 0.54 & 1  & 0.27 $\pm$ 0.30 & 1 &  &   \\
SR32  & 1.85 $\pm$ 0.39 & 5  & 0.79 $\pm$ 0.33 & 1 &  &   \\
SR33  & 1.20 $\pm$ 0.32 & 0  & 0.53 $\pm$ 0.13 & 0 &  &   \\
SR34  & 1.81 $\pm$ 0.42 & 3  & 6.3 $\pm$ 1.3 & 6 &  &   \\
SR35  & 1.98 $\pm$ 0.61 & 1  & 2.92 $\pm$ 0.87 & 3 &  &   \\
SR36  & 1.43 $\pm$ 0.37 & 2  & 0.51 $\pm$ 0.15 & 3 &  &   \\
SR37  & 4.2 $\pm$ 1.3 & 2  & 0.15 $\pm$ 0.07 & 0 &  &   \\
SR38  & 3.04 $\pm$ 0.68 & 4  & 1.07 $\pm$ 0.33 & 3 &  &   \\
SR39  & 0.63 $\pm$ 0.17 & 1  & 0.81 $\pm$ 0.47 & 0 &  &   \\
SR40  & 0.29 $\pm$ 0.34 & 0  & 1.54 $\pm$ 0.50 & 4 &  &   \\
SR41  & 0.80 $\pm$ 0.22 & 3  & 1.23 $\pm$ 0.53 & 1 &  &   \\
SR42  & 13.4 $\pm$ 1.9 & 19  & & &  &  \\
SR43  & 8.0 $\pm$ 3.0 & 8  & & &  &  \\
SR44  & 3.33 $\pm$ 0.74 & 3  & & &  &  \\
SR45  & 0.94 $\pm$ 0.26 & 1  & & &  &  \\
SR46  & 2.92 $\pm$ 0.50 & 3  & & &  &  \\
SR47  & 1.78 $\pm$ 0.42 & 3  & & &  &  \\
SR48  & 1.95 $\pm$ 0.39 & 5  & & &  &  \\
SR49  & 1.23 $\pm$ 0.30 & 3  & & &  &  \\
SR50  & 1.46 $\pm$ 0.31 & 0  & & &  &  \\
SR51  & 0.74 $\pm$ 0.18 & 0  & & &  &  \\
\hline
\end{tabular}
\end{table*}

\begin{figure*}[!hbtp]
\centering
\includegraphics[width=0.45\textwidth]{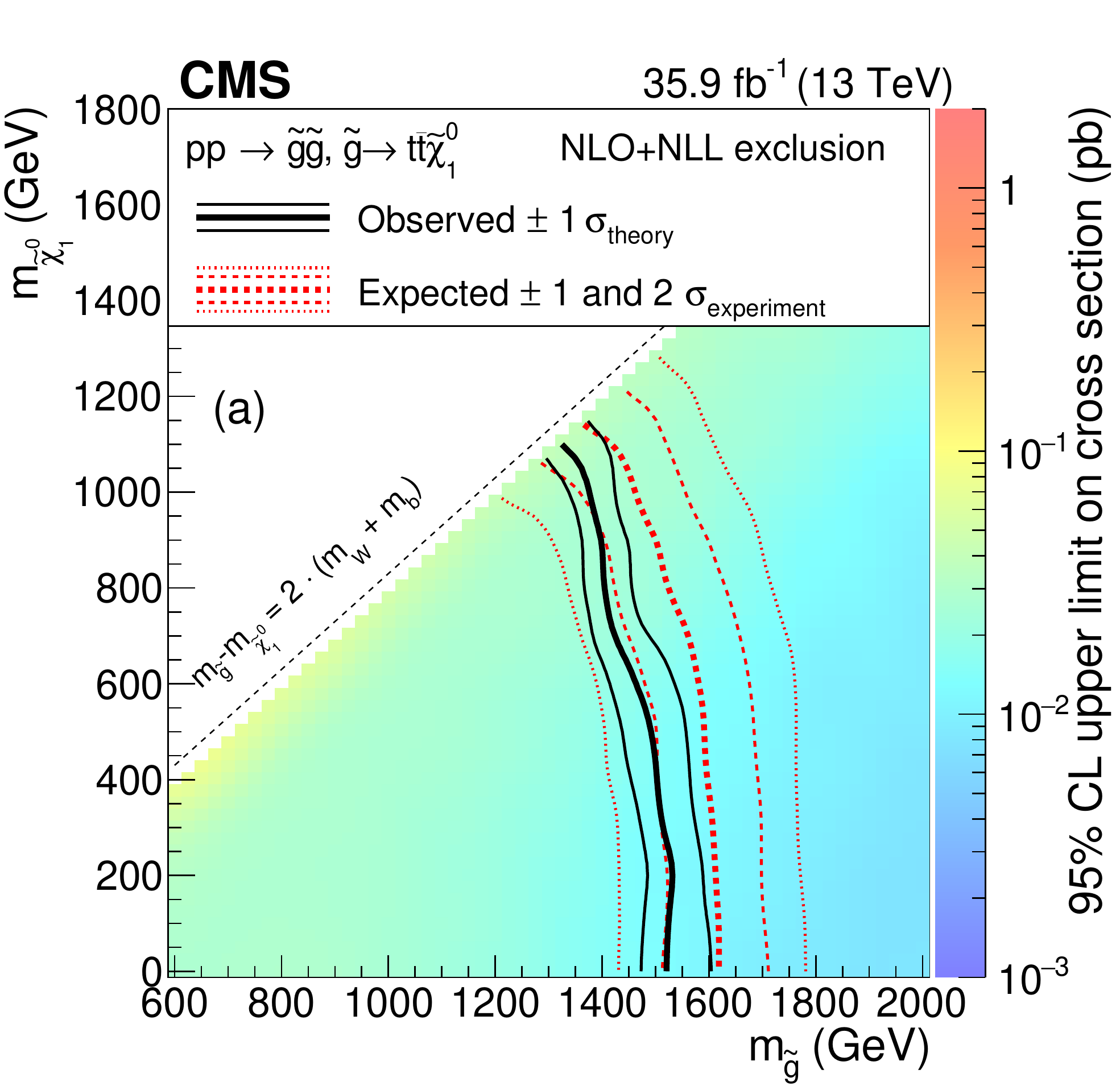}
\includegraphics[width=0.45\textwidth]{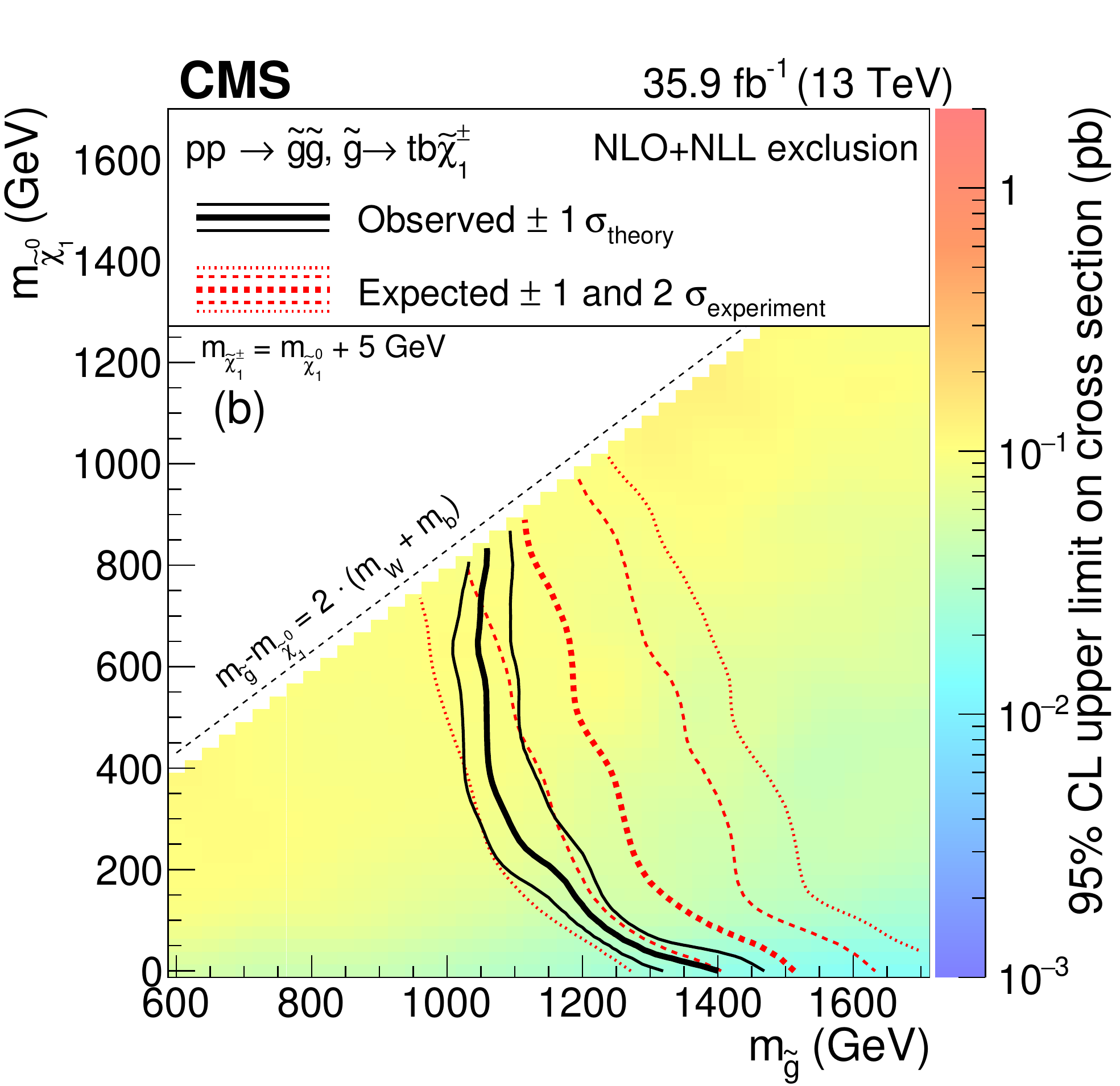}
\includegraphics[width=0.45\textwidth]{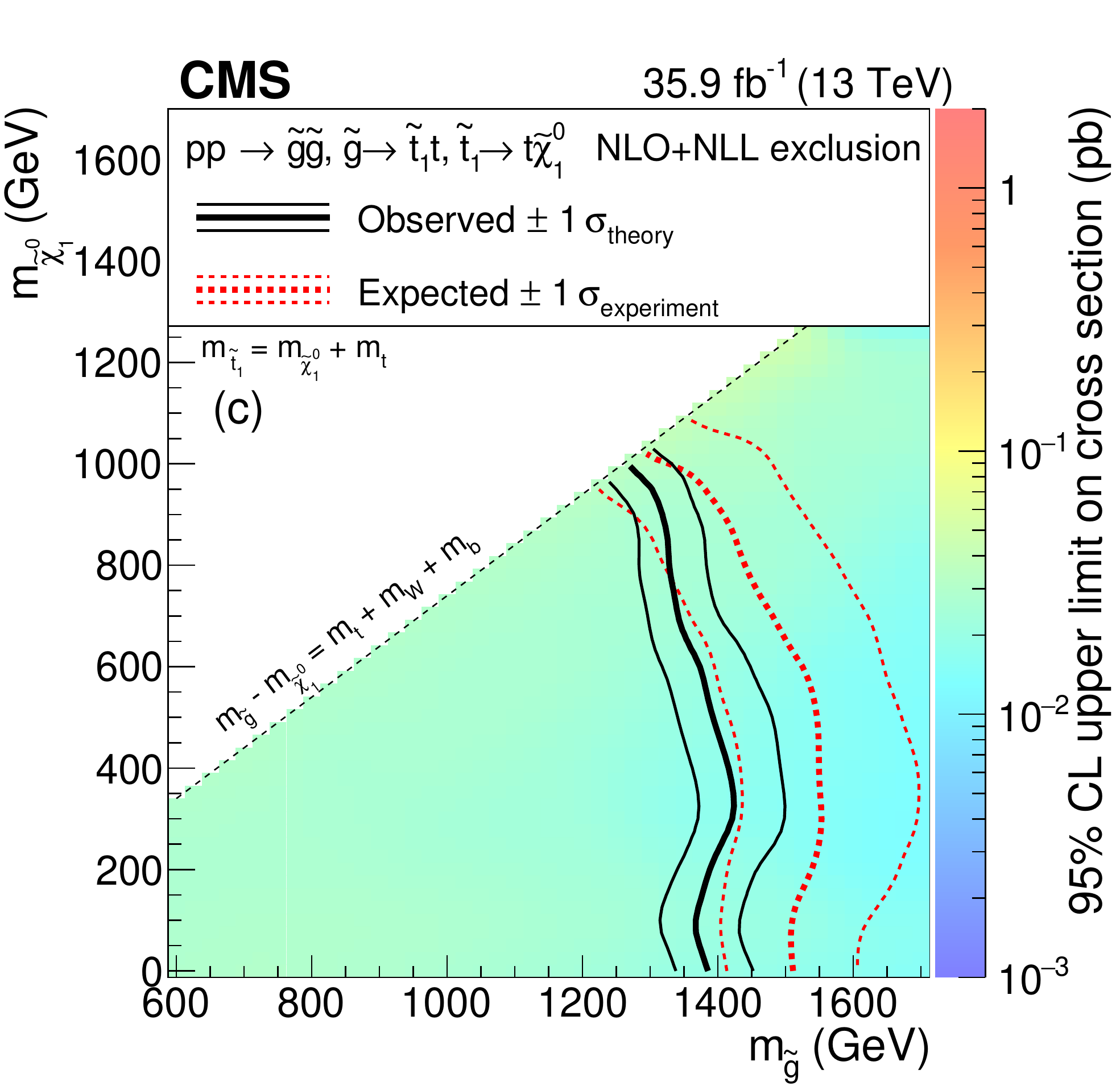}
\includegraphics[width=0.45\textwidth]{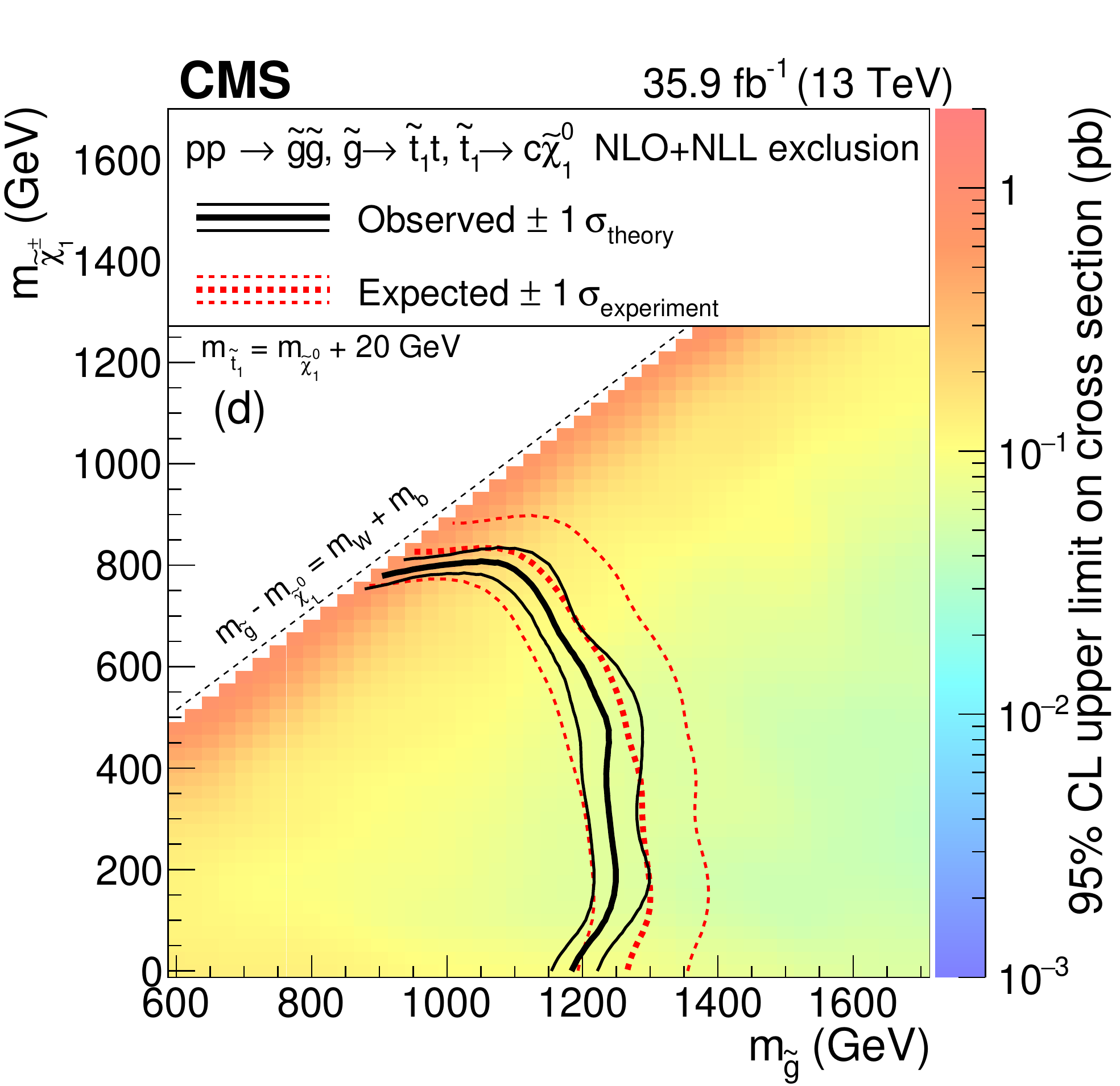}
\caption{ Exclusion regions at 95\% CL in the $m_{\lsp}$ versus
  $m_{\gluino}$ plane for the \Totttt~(a) and \TfttbbWW~(b) models, with off-shell third-generation squarks, and the
\Tftttt~(c) and \Tfttcc~(d) models, with on-shell third-generation squarks.
For the \TfttbbWW model, $m_{\chiplmin} = m_{\lsp} + 5\GeV$, for the \Tftttt model, $m_{\susytop} - m_{\lsp} = m_{\PQt}$, and
for the \Tfttcc model, $m_{\susytop} - m_{\lsp} = 20\GeV$ and the decay proceeds through $\susytop \to \PQc \lsp$.
The right-hand side color scale indicates the excluded cross section values for a given point in the SUSY particle mass plane.
The solid, black curves represent the observed exclusion limits
assuming the NLO+NLL cross
sections~\protect\cite{bib-nlo-nll-01,bib-nlo-nll-02,bib-nlo-nll-03,bib-nlo-nll-04,bib-nlo-nll-05,Borschensky:2014cia} (thick line), or their variations of $\pm$1 standard deviation (thin lines).
The dashed, red curves show the expected limits with the corresponding $\pm$1 and $\pm$2 standard deviation experimental uncertainties.
Excluded regions are to the left and below the limit curves.
}
\label{fig:t1ttxx_scan_xsec}
\end{figure*}

\begin{figure}[!hbtp]
\centering
\includegraphics[width=0.45\textwidth]{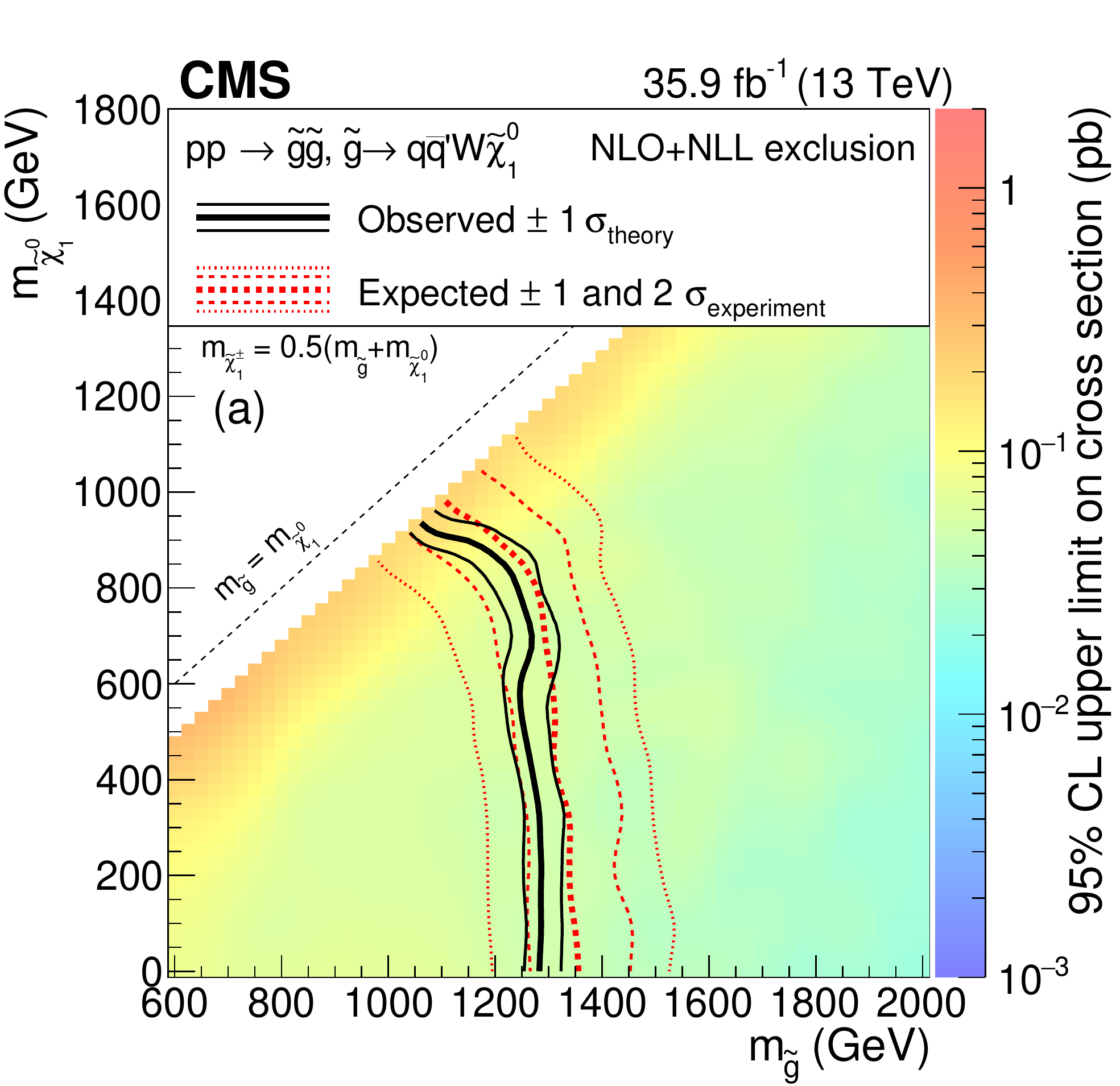}
\includegraphics[width=0.45\textwidth]{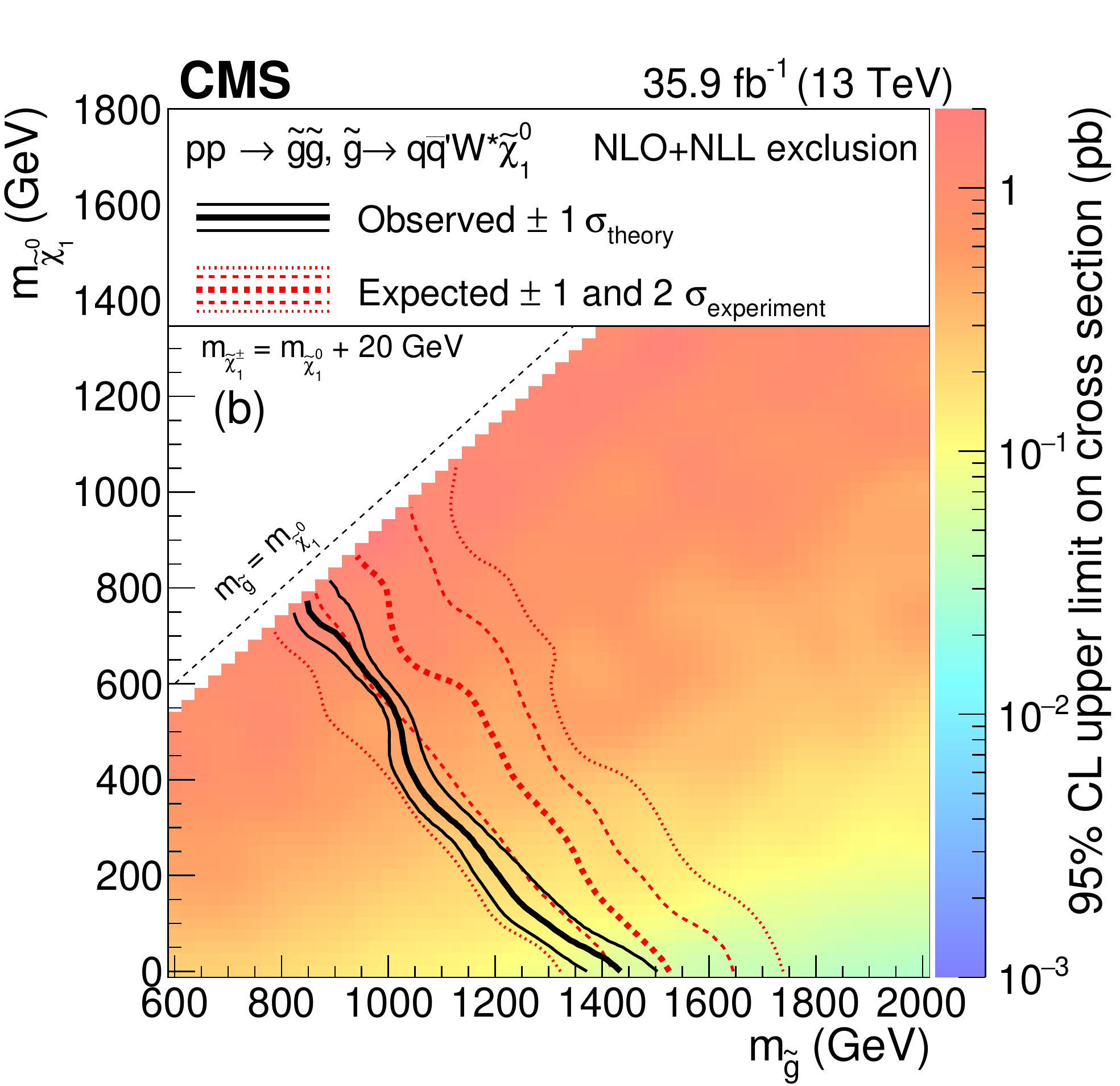}
\caption{Exclusion regions at 95\% CL in the plane of $m_{\lsp}$ versus $m_{\gluino}$ for the \TfqqqqWW model
with $m_{\chiplmin}=0.5(m_{\gluino} + m_{\lsp})$~(a) and with $m_{\chiplmin} = m_{\lsp} + 20\GeV$~(b).
The notations are as in Fig.~\protect\ref{fig:t1ttxx_scan_xsec}.}
\label{fig:t5qqqqww_scan_xsec}
\end{figure}

\begin{figure}[!hbtp]
\centering
\includegraphics[width=0.45\textwidth]{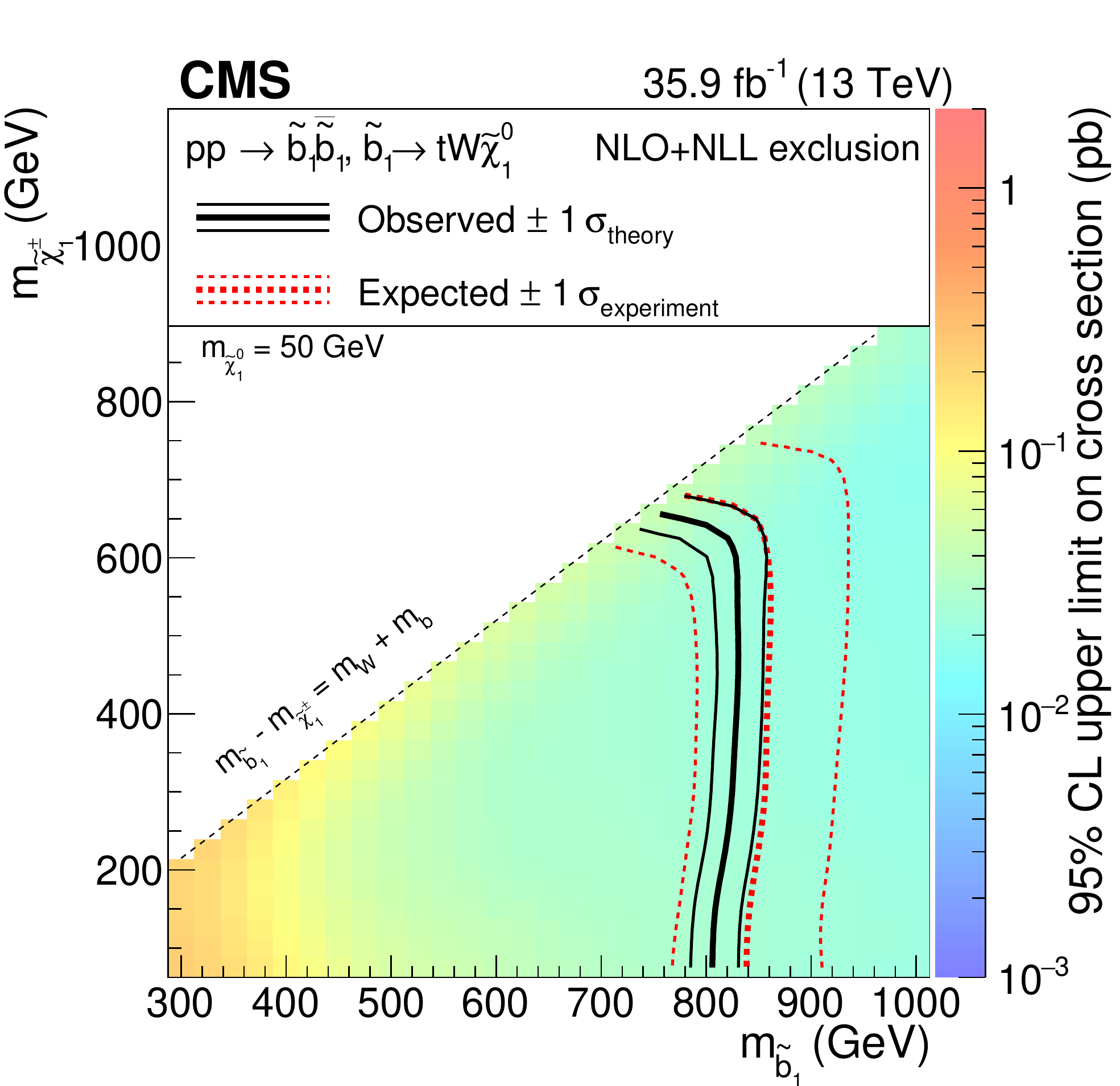}\\
\caption{ Exclusion regions at 95\% CL in the plane of $m_{\chiplmin}$ versus $m_{\sbottom}$ for the \TsttWW model with $m_{\lsp}=50\GeV$.
The notations are as in Fig.~\ref{fig:t1ttxx_scan_xsec}.}
\label{fig:t6ttww_scan_xsec}
\end{figure}

The observed SUSY cross section limits as a function of the gluino and LSP masses,
as well as the observed and expected mass limits for each simplified model,
are shown in Fig.~\ref{fig:t1ttxx_scan_xsec} for gluino pair production models with each gluino decaying through a
chain containing off- or on-shell third-generation squarks.
These models, which result in signatures with two or more \PQb quarks and two or more \PW\ bosons in the final state,
are introduced in Section~\ref{sec:samples} as \Totttt, \TfttbbWW, \Tftttt, and \Tfttcc.
Figure~\ref{fig:t5qqqqww_scan_xsec} shows the limits for a model of gluino production
followed by a decay through off-shell first- or second-generation squarks and a chargino.
Two different assumptions are made on the chargino mass, taken to be between that of
the gluino and the LSP.
These \TfqqqqWW models result in no \PQb quarks and either on-shell or off-shell \PW\ bosons.
Bottom squark pair production followed by a decay through a chargino, \TsttWW,
resulting in two \PQb quarks and four \PW\ bosons, is shown in Fig.~\ref{fig:t6ttww_scan_xsec}.
For all of the  models probed, the observed limit agrees well with the expected one,
extending the reach of the previous analysis by 200--300\GeV and reaching
1.5, 1.1, and 0.83\TeV for gluino, LSP, and bottom squark masses, respectively.

\begin{figure}
\centering
\includegraphics[width=0.45\textwidth]{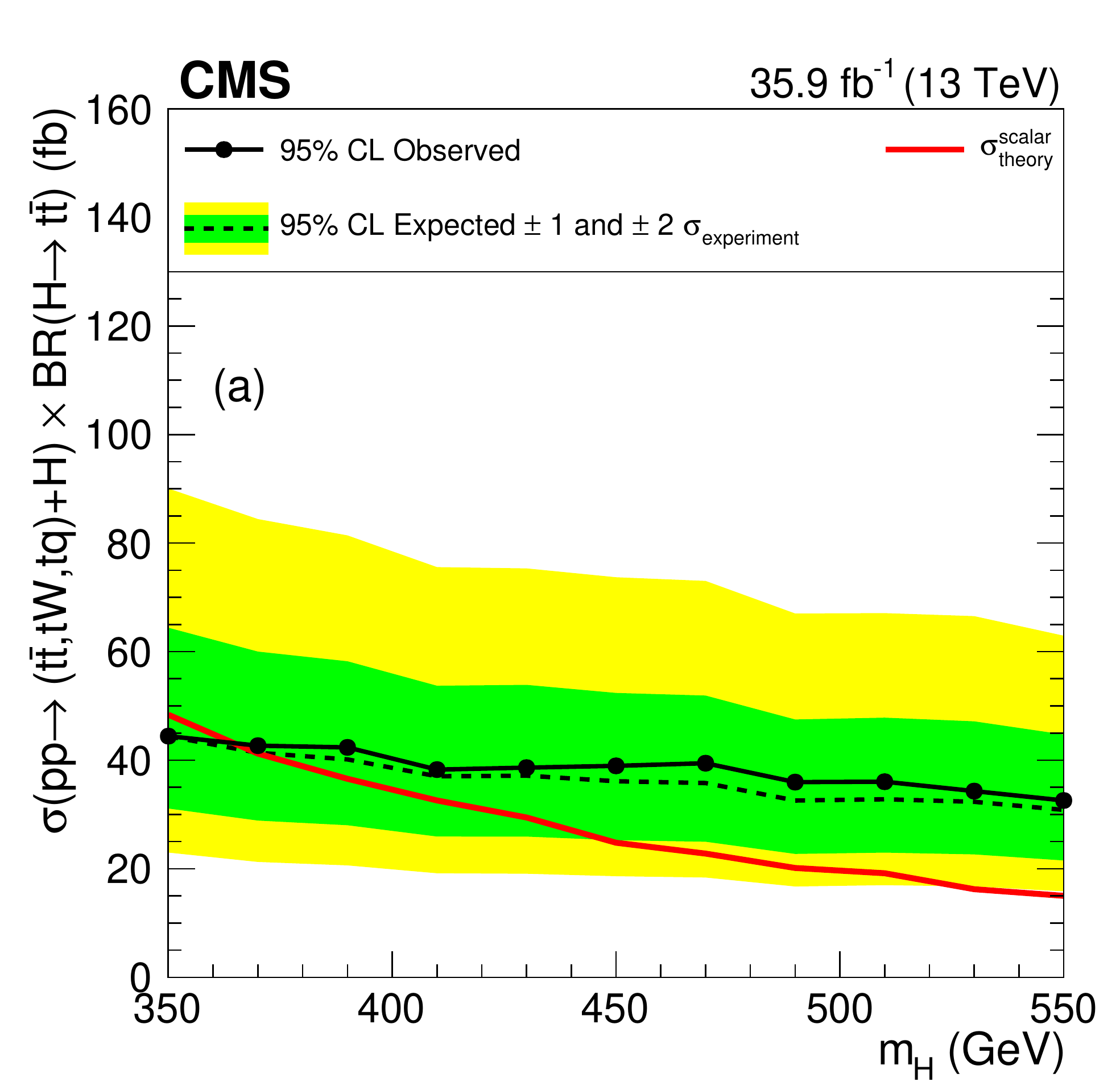}
\includegraphics[width=0.45\textwidth]{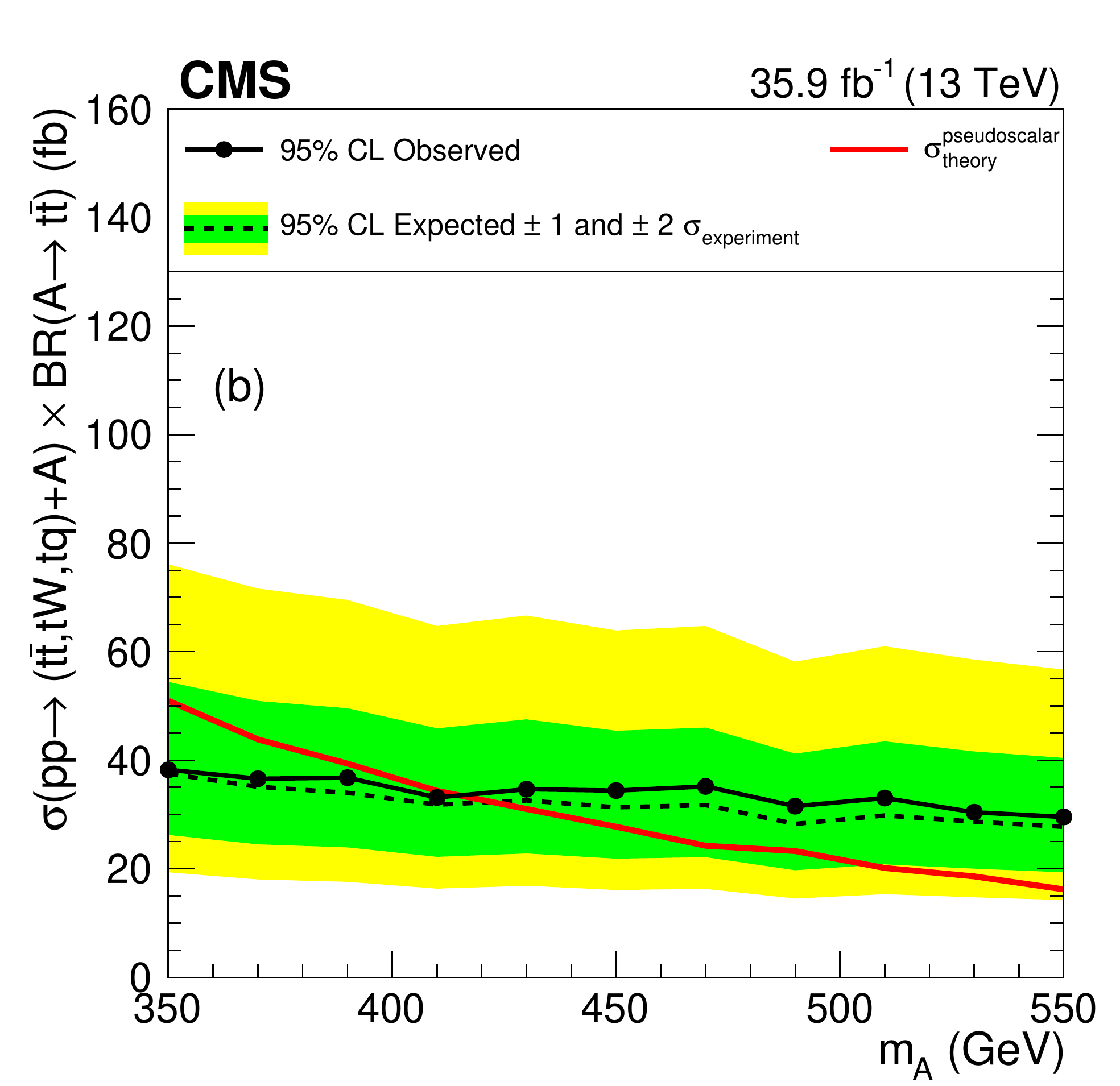}
\caption{Limits at 95\% CL on the production cross section for heavy scalar (a) and pseudoscalar (b)
boson in association to one or two top quarks, followed by its decay to top quarks, as a function of the (pseudo)scalar mass.
The red line corresponds to the theoretical cross section in the (pseudo)scalar model.
}
\label{fig:HiggsLimits}
\end{figure}

The observed and expected cross section limits on the production of a heavy scalar or a pseudoscalar boson
in association with one or two top quarks, followed by its decay to top quarks, are shown in Fig.~\ref{fig:HiggsLimits}.
The limits are compared with the total cross section of the processes described in Section~\ref{sec:samples}.
The observed limit, which agrees well with the expected one,
excludes scalar (pseudoscalar) masses up to 360~(410)\GeV.

{\tolerance=1200
The SM four top quark production,
$\Pp \Pp \to \ttbar \ttbar$,
is normally included among the rare SM backgrounds.
When treating this process as signal, its observed (expected) cross
section limit is determined to be 42 ($27^{+13}_{-8}$)\unit{fb} at 95\% CL, to be compared to the SM expectation of $9.2^{+2.9}_{-2.4}$\unit{fb}\cite{MADGRAPH5}.
This is a significant improvement with respect to the observed (expected) limits obtained in the previous version of this analysis,
 119 ($102^{+57}_{-35}$)\unit{fb}\cite{SUS-15-008}, as well as the combination of those results with results from
 single-lepton and opposite-sign dilepton final states, 69 ($71^{+38}_{-24}$)\unit{fb}~\cite{CMStttt2015}.
\par}

The results of the search are also used to set a limit on the
production cross section for SS top quark pairs,
$\sigma(\Pp\Pp \to \PQt\PQt) + \sigma(\Pp\Pp \to \cPaqt\cPaqt)$.
The observed (expected) limit, based on the kinematics of a SM \ttbar sample
and determined using the number of b jets distribution
in the baseline region, is 1.2 ($0.76^{+0.3}_{-0.2}$)\unit{pb} at 95\% CL, significantly improved with respect to the
1.7 ($1.5^{+0.7}_{-0.4}$)\unit{pb} observed (expected) limit of the previous analysis~\cite{SUS-15-008}.

\subsection{Model-independent limits and additional results}
\label{sec:aggregate}

The yields and background predictions can be used
to test additional BSM physics scenarios. To facilitate such reinterpretations,
we provide limits on the number of SS dilepton pairs as a function of the \MET and \HT thresholds in the kinematic tails,
as well as results from a smaller number of inclusive and exclusive  signal regions.

\begin{figure}
\centering
\includegraphics[width=0.45\textwidth]{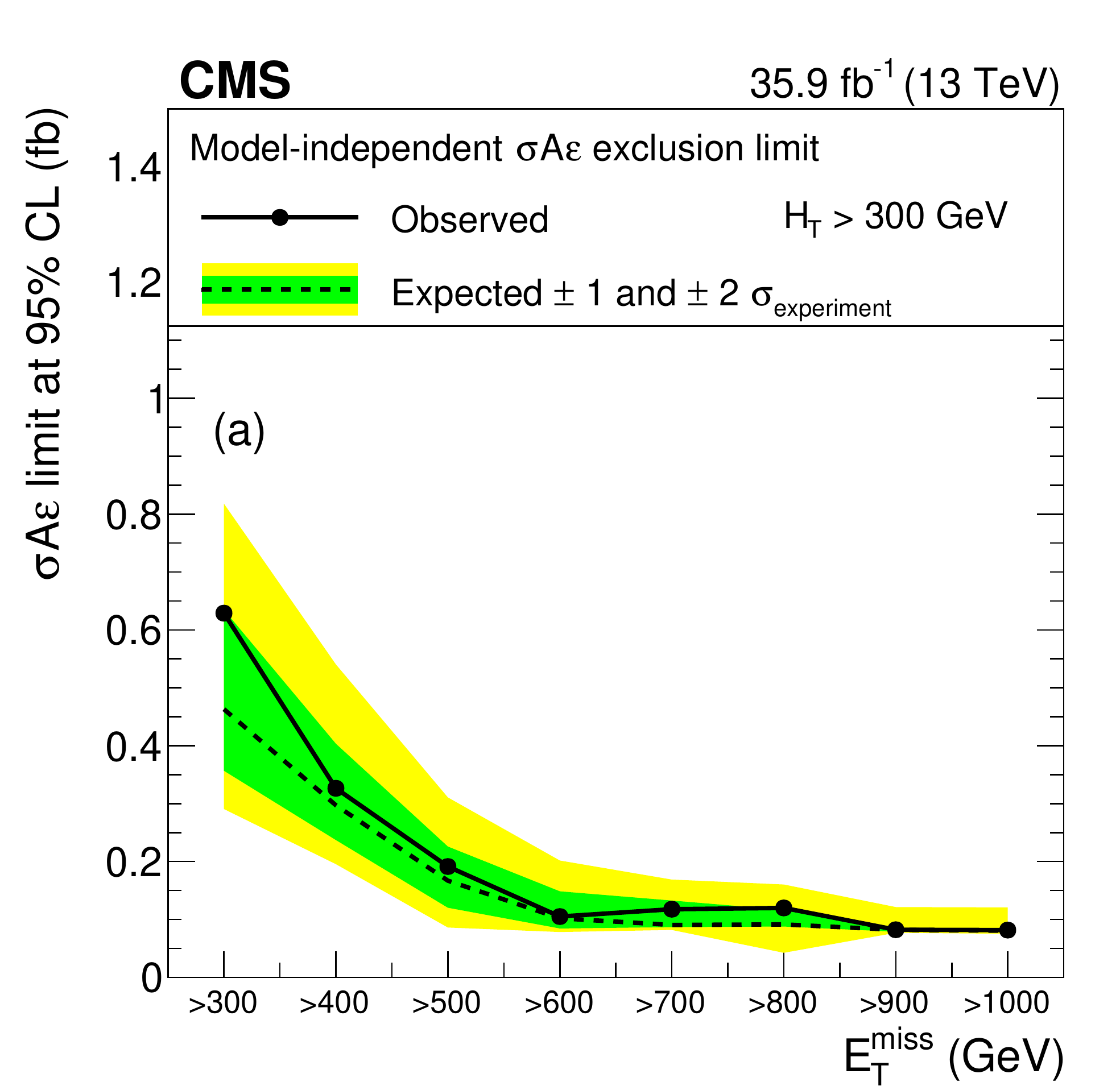}
\includegraphics[width=0.45\textwidth]{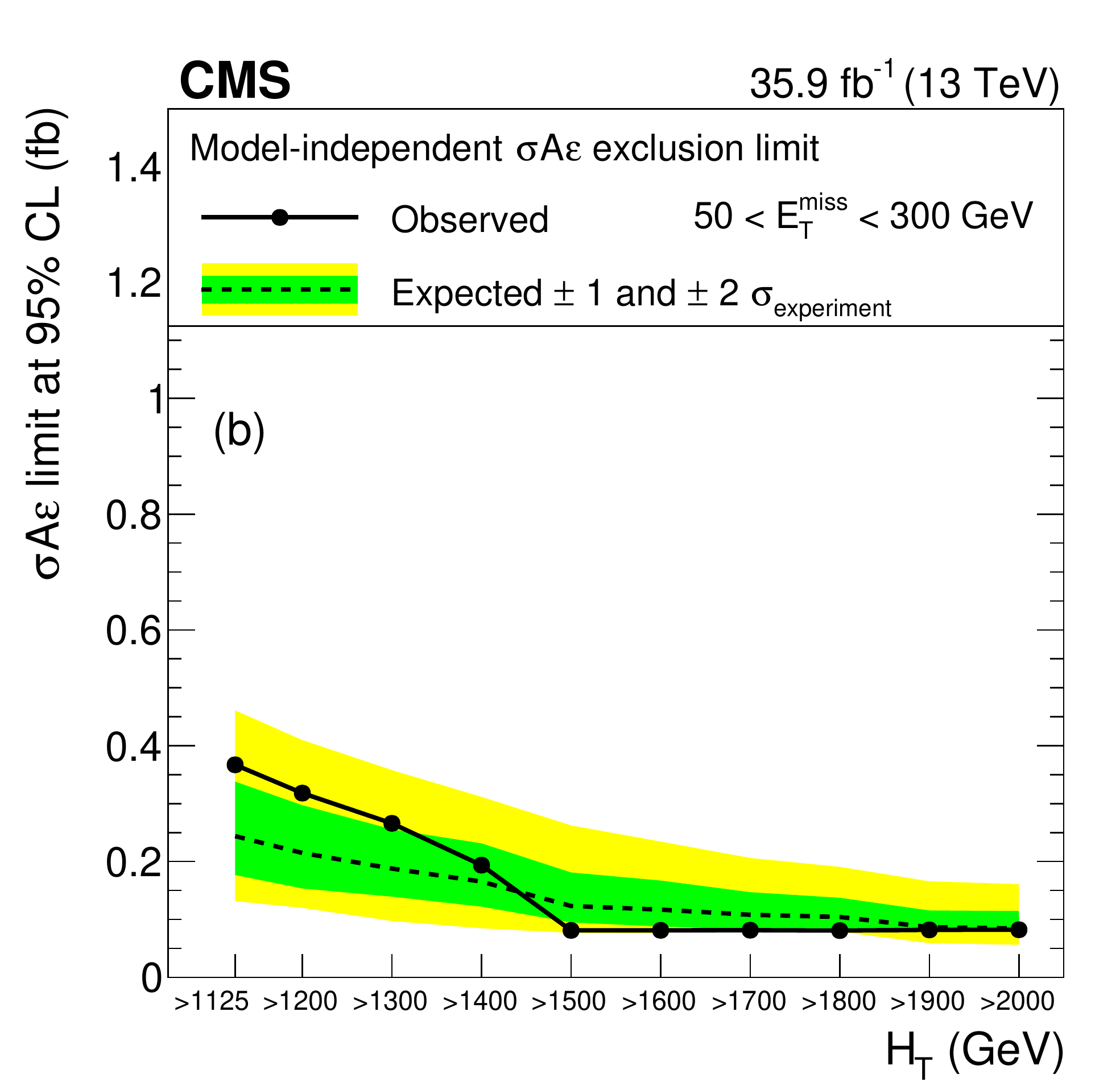}
\caption{Limits on the product of cross section, detector acceptance, and selection efficiency, $\sigma \! \mathcal{A} \epsilon$,
for the production of an SS dilepton pair as a function of the
\MET~(a) and of \HT~(b) thresholds.
}
\label{fig:MIlimits}
\end{figure}

The \MET and \HT limits are based on combining HH tail SRs,
specifically SR42--45 for high \MET and SR46--51 for high \HT,
and employing the CL$_\mathrm{s}$ criterion without the asymptotic formulation
as a function of the minimum threshold of each kinematic variable.
These limits are presented in Fig.~\ref{fig:MIlimits} in terms of $\sigma \! \mathcal{A} \epsilon$,
the product of cross section, detector acceptance, and selection efficiency.
Where no events are observed, the observed and expected limits reach 0.1\unit{fb},
to be compared with a limit of 1.3\unit{fb} obtained in the previous analysis~\cite{SUS-15-008}.

Results are also provided in Table~\ref{tab:inclusive_aggregate_def} for
a small number of inclusive signal regions,
designed based on different topologies and
a small number of expected background events.
The background expectation, the event count, and the expected BSM yield in any one of these regions can be used to constrain BSM hypotheses in a simple way.

\begin{figure*}
\centering
\includegraphics[width=\textwidth]{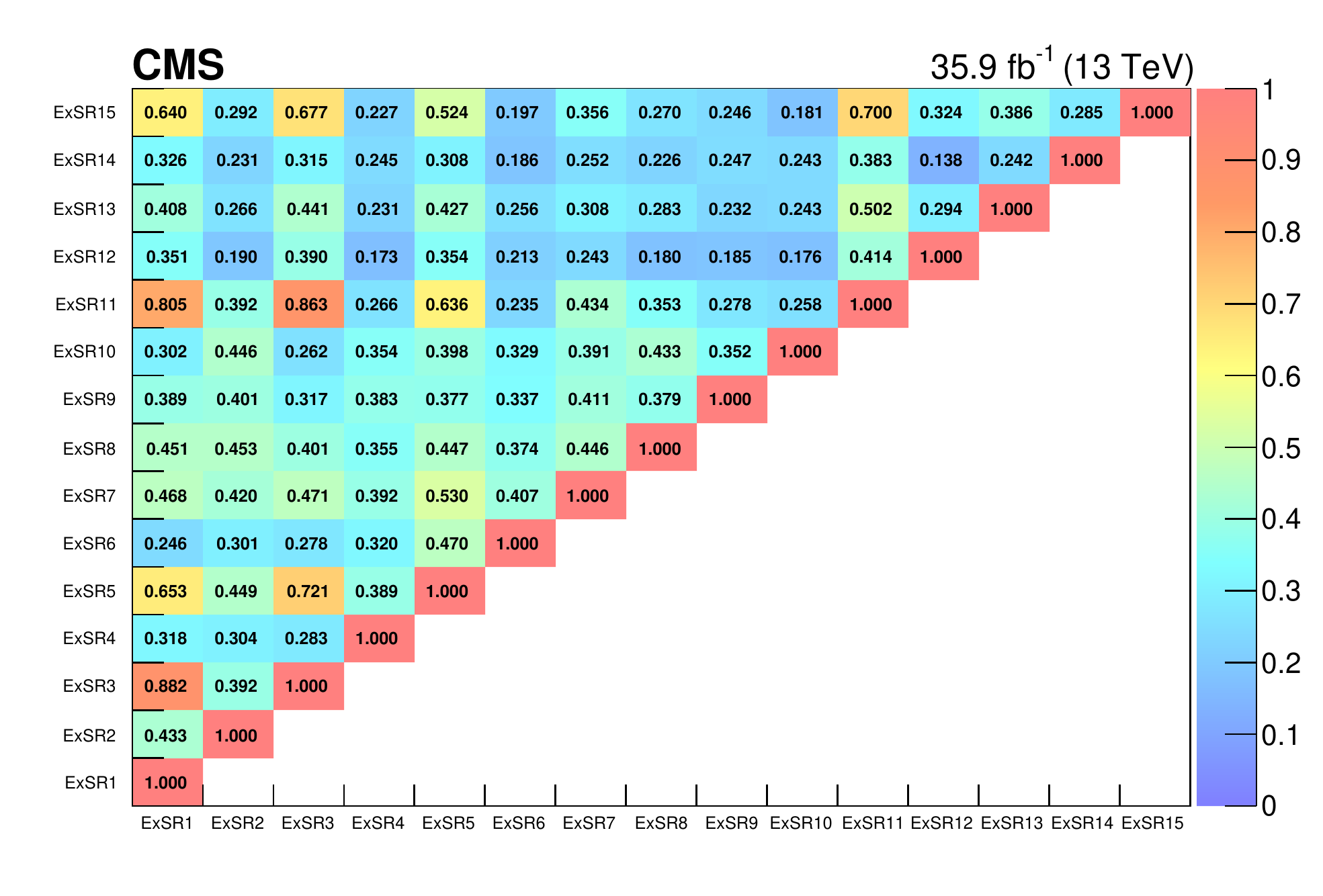}
 \caption{Correlations between the background predictions in the 15 exclusive regions. }
\label{fig:correlation_exclusive}
\end{figure*}

In addition, we define a small number of exclusive signal regions based on integrating over the standard signal regions.
Their definitions, as well as the expected and observed yields, are specified in Table~\ref{tab:exclusive_aggregate},
while the correlation matrix for the background predictions in these regions is given in Fig.~\ref{fig:correlation_exclusive}.  This
information can be used to construct a simplified likelihood for models of new physics, as described in Ref.~\cite{Collaboration:2242860}.

\begin{table*}
\centering
\topcaption{Inclusive SR definitions, expected background yields, and observed yields, as well the observed 95\% CL upper limits on the number of signal events contributing to each region.
No uncertainty in the signal acceptance is assumed in calculating these limits. A dash (---) means that the selection is not applied. }
\label{tab:inclusive_aggregate_def}
\cmsTable{
  \begin{tabular}{|c|c|c|c|c|c|c|c|c|c|}
  \hline
      \rule{0pt}{10pt}
         SR     & Leptons              & \Njets   & \Nbjets  & \HT (\GeVns{}) & \MET (\GeVns{})& \MTmin (\GeVns{})& SM expected       & Observed  &  $N_\text{obs,UL}^\mathrm{95\% CL}$ \\
         \hline\hline
        InSR1    & \multirow{11}{*}{HH} & $\geq$2 & 0        & $\geq$1200 & $\geq$50   & \NA             &   4.00 $\pm$   0.79 &     10 & 12.35 \\
        InSR2    &                      & $\geq$2 & $\geq$2 & $\geq$1100 & $\geq$50   & \NA             &   3.63 $\pm$   0.71 &      4 & 5.64  \\
        InSR3    &                      & $\geq$2 & 0        & \NA           & $\geq$450  & \NA             &   3.72 $\pm$   0.83 &      4 & 5.62  \\
        InSR4    &                      & $\geq$2 & $\geq$2 & \NA           & $\geq$300  & \NA             &   3.32 $\pm$   0.81 &      6 & 8.08  \\
        InSR5    &                      & $\geq$2 & 0        & \NA           & $\geq$250  & $\geq$120    &   1.68 $\pm$   0.44 &      2 & 4.46  \\
        InSR6    &                      & $\geq$2 & $\geq$2 & \NA           & $\geq$150  & $\geq$120    &   3.82 $\pm$   0.76 &      7 & 9.06  \\
        InSR7    &                      & $\geq$2 & 0        & $\geq$900  & $\geq$200  & \NA             &    5.6 $\pm$    1.1 &     10 & 10.98 \\
        InSR8    &                      & $\geq$2 & $\geq$2 & $\geq$900  & $\geq$200  & \NA             &    5.8 $\pm$    1.3 &      9 & 9.77  \\
        InSR9    &                      & $\geq$7 & \NA        & \NA           & $\geq$50   & \NA             &   10.1 $\pm$    2.7 &      9 & 7.39  \\
        InSR10   &                      & $\geq$4 & \NA        & \NA           & $\geq$50   & $\geq$120    &   15.2 $\pm$    3.5 &     22 & 16.73 \\
        InSR11   &                      & $\geq$2 & $\geq$3 & \NA           & $\geq$50   & \NA             &   13.3 $\pm$    3.4 &     17 & 13.63 \\
        InSR12   & \multirow{4}{*}{LL}  & $\geq$2 & 0        & $\geq$700  & $\geq$50   & \NA             &    3.6 $\pm$    2.5 &      3 & 4.91  \\  \hline
        InSR13   &                      & $\geq$2 & \NA        & \NA           & $\geq$200  & \NA             &    4.9 $\pm$    2.9 &     10 & 11.76 \\
        InSR14   &                      & $\geq$5 & \NA        & \NA           & $\geq$50   & \NA             &    7.3 $\pm$    5.5 &      6 & 6.37  \\
        InSR15   &                      & $\geq$2 & $\geq$3 & \NA           & $\geq$50   & \NA             &   1.06 $\pm$   0.99 &      0 & 2.31  \\   \hline
\end{tabular}
}
\end{table*}

\begin{table*}
\centering
\topcaption{Exclusive SR definitions, expected background yields, and observed yields. A dash (---) means that the selection is not applied.}
\label{tab:exclusive_aggregate}
\cmsTable{
  \begin{tabular}{|c|c|c|c|c|c|c|c|c|}
  \hline
         SR      & Leptons               & \Njets    & \Nbjets   & \MET (\GeVns{}) & \HT (\GeVns{})  & \MTmin (\GeVns{})         & SM expected          & Observed  \\
         \hline\hline
        ExSR1  & \multirow{11}{*}{HH} & $\geq$2 & 0        & 50--300    & $<$1125     & $<$120 for $\HT>300$ &    700 $\pm$ 130 & 685  \\
        ExSR2  &                      & $\geq$2 & 0        & 50--300    & 300--1125   & $\geq$120            &       11.0 $\pm$ 2.2 & 11  \\
        ExSR3  &                      & $\geq$2 & 1        & 50--300    & $<$1125    & $<$120 for $\HT>300$ &    477 $\pm$ 120 & 482  \\
        ExSR4  &                      & $\geq$2 & 1        & 50--300    & 300--1125   & $\geq$120            &        8.4 $\pm$ 3.5 & 8  \\
        ExSR5  &                      & $\geq$2 & 2        & 50--300    & $<$1125    & $<$120 for $\HT>300$ &     137 $\pm$ 25 & 152  \\
        ExSR6  &                      & $\geq$2 & 2        & 50--300    & 300--1125   & $\geq$120            &        4.9 $\pm$ 1.2 & 8  \\
        ExSR7  &                      & $\geq$2 & $\geq$3 & 50--300    & $<$1125    & $<$120 for $\HT>300$ &       11.6 $\pm$ 3.1 & 10  \\
        ExSR8  &                      & $\geq$2 & $\geq$3 & 50--300    & 300--1125   & $\geq$120            &       0.8 $\pm$ 0.24 & 3  \\
        ExSR9  &                      & $\geq$2 & \NA        & $\geq$300 & $\geq$300  & \NA                     &       25.7 $\pm$ 5.4 & 31  \\
        ExSR10 &                      & $\geq$2 & \NA        & 50--300    & $\geq$1125 & \NA                     &       10.1 $\pm$ 2.2 & 14  \\ \hline
        ExSR11 & \multirow{4}{*}{HL}  & $\geq$2 & \NA        & 50--300    & $<$1125    & $<$120               &   1070 $\pm$ 250 & 1167  \\
        ExSR12 &                      & $\geq$2 & \NA        & 50--300    & $<$1125    & $\geq$120            &      1.33 $\pm$ 0.46 & 1  \\
        ExSR13 &                      & $\geq$2 & \NA        & $\geq$300 & $\geq$300  & \NA                     &        9.9 $\pm$ 2.5 & 12  \\
        ExSR14 &                      & $\geq$2 & \NA        & 50--300    & $\geq$1125 & \NA                     &        4.7 $\pm$ 1.8 & 8  \\  \hline
        ExSR15 & LL                   & $\geq$2 & \NA        & $\geq$50  & $\geq$300  & \NA                     &      37 $\pm$ 12 & 43  \\    \hline
\end{tabular}
}
\end{table*}

\section{Summary}
\label{sec:summary}
A sample of same-sign dilepton events produced in proton-proton collisions at 13\TeV,
corresponding to an integrated luminosity of \sslumi, has been studied to search for
manifestations of physics beyond the standard model.
The data are found to be consistent with the standard model expectations, and no excess event yield is observed.
The results are interpreted as limits at 95\% confidence level on  cross sections for the
production of new particles in simplified supersymmetric models.
Using calculations for these cross sections as functions of particle
masses, the limits are turned into lower mass limits
that are as high as 1500\GeV for gluinos and 830\GeV for bottom squarks,
depending on the details of the model.
Limits are also provided on the production of heavy
scalar (excluding the mass range 350--360\GeV) and pseudoscalar (350--410\GeV) bosons
decaying to top quarks in the context of two Higgs doublet models,
as well as on same-sign top quark pair production,
and the standard model production of four top quarks.
Finally, to facilitate further interpretations of the search, model-independent
limits are provided as a function of \HT and \MET,
together with the background prediction and data yields in a smaller set of
signal regions.

\begin{acknowledgments}
We congratulate our colleagues in the CERN accelerator departments for the excellent performance of the LHC and thank the technical and administrative staffs at CERN and at other CMS institutes for their contributions to the success of the CMS effort. In addition, we gratefully acknowledge the computing centers and personnel of the Worldwide LHC Computing Grid for delivering so effectively the computing infrastructure essential to our analyses. Finally, we acknowledge the enduring support for the construction and operation of the LHC and the CMS detector provided by the following funding agencies: BMWFW and FWF (Austria); FNRS and FWO (Belgium); CNPq, CAPES, FAPERJ, and FAPESP (Brazil); MES (Bulgaria); CERN; CAS, MoST, and NSFC (China); COLCIENCIAS (Colombia); MSES and CSF (Croatia); RPF (Cyprus); SENESCYT (Ecuador); MoER, ERC IUT, and ERDF (Estonia); Academy of Finland, MEC, and HIP (Finland); CEA and CNRS/IN2P3 (France); BMBF, DFG, and HGF (Germany); GSRT (Greece); OTKA and NIH (Hungary); DAE and DST (India); IPM (Iran); SFI (Ireland); INFN (Italy); MSIP and NRF (Republic of Korea); LAS (Lithuania); MOE and UM (Malaysia); BUAP, CINVESTAV, CONACYT, LNS, SEP, and UASLP-FAI (Mexico); MBIE (New Zealand); PAEC (Pakistan); MSHE and NSC (Poland); FCT (Portugal); JINR (Dubna); MON, RosAtom, RAS, RFBR and RAEP (Russia); MESTD (Serbia); SEIDI, CPAN, PCTI and FEDER (Spain); Swiss Funding Agencies (Switzerland); MST (Taipei); ThEPCenter, IPST, STAR, and NSTDA (Thailand); TUBITAK and TAEK (Turkey); NASU and SFFR (Ukraine); STFC (United Kingdom); DOE and NSF (USA).

\hyphenation{Rachada-pisek} Individuals have received support from the Marie-Curie program and the European Research Council and EPLANET (European Union); the Leventis Foundation; the A. P. Sloan Foundation; the Alexander von Humboldt Foundation; the Belgian Federal Science Policy Office; the Fonds pour la Formation \`a la Recherche dans l'Industrie et dans l'Agriculture (FRIA-Belgium); the Agentschap voor Innovatie door Wetenschap en Technologie (IWT-Belgium); the Ministry of Education, Youth and Sports (MEYS) of the Czech Republic; the Council of Science and Industrial Research, India; the HOMING PLUS program of the Foundation for Polish Science, cofinanced from European Union, Regional Development Fund, the Mobility Plus program of the Ministry of Science and Higher Education, the National Science Center (Poland), contracts Harmonia 2014/14/M/ST2/00428, Opus 2014/13/B/ST2/02543, 2014/15/B/ST2/03998, and 2015/19/B/ST2/02861, Sonata-bis 2012/07/E/ST2/01406; the National Priorities Research Program by Qatar National Research Fund; the Programa Clar\'in-COFUND del Principado de Asturias; the Thalis and Aristeia programs cofinanced by EU-ESF and the Greek NSRF; the Rachadapisek Sompot Fund for Postdoctoral Fellowship, Chulalongkorn University and the Chulalongkorn Academic into Its 2nd Century Project Advancement Project (Thailand); and the Welch Foundation, contract C-1845. \end{acknowledgments}
\clearpage
\bibliography{auto_generated}

\cleardoublepage \appendix\section{The CMS Collaboration \label{app:collab}}\begin{sloppypar}\hyphenpenalty=5000\widowpenalty=500\clubpenalty=5000\input{SUS-16-035-authorlist.tex}\end{sloppypar}
\end{document}

%% file: SUS-16-035-authorlist.tex
\textbf{Yerevan Physics Institute,  Yerevan,  Armenia}\\*[0pt]
A.M.~Sirunyan, A.~Tumasyan
\vskip\cmsinstskip
\textbf{Institut f\"{u}r Hochenergiephysik,  Wien,  Austria}\\*[0pt]
W.~Adam, F.~Ambrogi, E.~Asilar, T.~Bergauer, J.~Brandstetter, E.~Brondolin, M.~Dragicevic, J.~Er\"{o}, M.~Flechl, M.~Friedl, R.~Fr\"{u}hwirth\cmsAuthorMark{1}, V.M.~Ghete, J.~Grossmann, J.~Hrubec, M.~Jeitler\cmsAuthorMark{1}, A.~K\"{o}nig, N.~Krammer, I.~Kr\"{a}tschmer, D.~Liko, T.~Madlener, I.~Mikulec, E.~Pree, D.~Rabady, N.~Rad, H.~Rohringer, J.~Schieck\cmsAuthorMark{1}, R.~Sch\"{o}fbeck, M.~Spanring, D.~Spitzbart, J.~Strauss, W.~Waltenberger, J.~Wittmann, C.-E.~Wulz\cmsAuthorMark{1}, M.~Zarucki
% ...
\vskip\cmsinstskip
% ...
1:~~Also at Vienna University of Technology, Vienna, Austria\\
% ...
